\documentclass[journal]{IEEEtran}
\usepackage[left=0.75in, right=0.75in, top=1in, bottom=0.9in]{geometry}
\usepackage{mathrsfs,amsthm}
\usepackage{cite}

\makeatletter
\def\ps@headings{%
\def\@oddhead{\mbox{}\scriptsize\rightmark \hfil \thepage}%
\def\@evenhead{\scriptsize\thepage \hfil \leftmark\mbox{}}%
\def\@oddfoot{}%
\def\@evenfoot{}}
\makeatother 
\usepackage{amsmath,amssymb,stfloats,subfigure,multicol}
\usepackage{float}
\usepackage{graphicx}
\usepackage{epsfig,epsf,psfrag,amssymb,amsfonts,latexsym,graphicx,mathrsfs,subfigure}
\usepackage{bbm}
\usepackage{dsfont}
\usepackage{chngpage}
\usepackage[dvips]{color}
\usepackage{textcomp}
\usepackage{hyperref}
\usepackage{url}
\usepackage[hyphenbreaks]{breakurl}
\usepackage[normalem]{ulem}
\usepackage{algpseudocode}
\usepackage{caption}
\newsavebox{\ieeealgbox}

\usepackage{array}

\newtheorem{proposition}{Proposition}

\newcommand*{\QED}{\hfill\ensuremath{\square}}

 \def\old#1{}    

\def\ED{{\mbox{\rm\tiny ED}}}
\def\RED{{\mbox{\rm\tiny R-ED}}}
\def\TLMP{{\mbox{\rm\tiny TLMP}}}
\def\RTLMP{{\mbox{\rm\tiny R-TLMP}}}
\def\LMP{{\mbox{\rm\tiny LMP}}}
\def\RLMP{{\mbox{\rm\tiny R-LMP}}}
\def\RPMP{{\mbox{\rm\tiny R-PMP}}}
\def\RCMP{{\mbox{\rm\tiny R-CMP}}}
\def\MLMP{{\mbox{\rm\tiny M-LMP}}}
\def\oneED{{\mbox{\rm\tiny 1-ED}}}

\def\nn{\nonumber}
\def\beq{\begin{equation}}
\def\eeq{\end{equation}}
\def\bea{\begin{eqnarray}}
\def\eea{\end{eqnarray}}
\def\ba{\begin{array}}
\def\ea{\end{array}}

\def\bitem{\begin{itemize}}
\def\eitem{\end{itemize}}
\def\ben{\begin{enumerate}}
\def\een{\end{enumerate}}

\def\etal{{\it et al. \/}}

\def\ie{{\it i.e.,\ \/}}

\definecolor{bgrd}{rgb}{1,1,1}
\definecolor{gray}{rgb}{0.5,0.5,0.5}
\definecolor{dkr}{rgb}{0.7,0.1,0.2}
\definecolor{dkb}{rgb}{0.1,0.1,0.8}

\makeatletter
\newdimen{\captionwidth}
\long\def\@makecaption#1#2{%
\captionwidth .9\hsize
\vskip 10pt%
\setbox\@tempboxa\hbox{#1: #2}%
  \ifdim \wd\@tempboxa >\captionwidth%
    \setbox\@tempboxa\hbox{#1:\hspace*{.5em}}%
    \hfil\parbox{\captionwidth}{\raggedright\hangindent \wd\@tempboxa%
    \hangafter=1\unhbox\@tempboxa#2}\hfill%
  \else\centerline{\box\@tempboxa}%
  \fi
}
\makeatother
\def\scalefig#1{\epsfxsize #1\textwidth}

\def\edoc{\end{document}}

\newcommand{\Gmsc}{\mathscr{G}}
\newcommand{\Hmsc}{\mathscr{H}}

\def\pibf{\hbox{\boldmath$\pi$\unboldmath}}

\def\cbf{{\pmb c}}
\def\dbf{{\pmb d}}

\def\gbf{{\pmb g}}

\def\pbf{{\pmb p}}
\def\qbf{{\pmb q}}

\def\sbf{{\pmb s}}

\def\zbf{{\pmb z}}

\def\Gbf{{\pmb G}}

\def\Sbf{{\pmb S}}

\def\Gc{{\cal G}}

\def\Pc{{\cal P}}

\begin{document}

\title{Pricing Multi-Interval Dispatch under Uncertainty\\
Part II: Generalization and Performance}
\author{Cong Chen,~\IEEEmembership{Student Member,~IEEE,}
Ye Guo,~\IEEEmembership{Senior Member,~IEEE,}
		and~Lang~Tong,~\IEEEmembership{Fellow,~IEEE}
\thanks{\scriptsize Part of the work was presented at the 56th Allerton Conference on Communication, Control, and Computing \cite{Guo&Tong:18Allerton} and 2019 IEEE PESGM \cite{Tong:19PESGM}.}
\thanks{\scriptsize
Cong Chen (\url{cc2662@cornell.edu}) and Lang Tong (\url{lt35@cornell.edu}) are with the Cornell University, Ithaca, NY 14853, USA. Ye Guo (\url{guo-ye@sz.tsinghua.edu.cn}) is with Tsinghua Berkeley Shenzhen Institute, Shenzhen, P.R. China. }
\thanks{\scriptsize The work of L. Tong and C. Chen is supported in part by the National Science Foundation under
Award 1809830 and 1932501, Power Systems and Engineering Research Center (PSERC) Research Project M-39. The work of Y. Guo is supported in part of the National Science Foundation of China under Award 51977115.Corresponding authors: Lang Tong and Ye Guo.}}
\maketitle

\begin{abstract}
Pricing multi-interval economic dispatch of electric power under operational uncertainty is considered in this two-part paper.  Part I investigates dispatch-following incentives for generators under the locational marginal pricing (LMP) and temporal locational marginal pricing (TLMP) policies.  Extending the theoretical results developed in Part I, Part II evaluates a broader set of performance measures under a general network model.  For networks with power flow constraints, TLMP is shown to have an energy-congestion-ramping price decomposition.  Under the one-shot dispatch and pricing model, this decomposition leads to a nonnegative merchandising surplus equal to the sum of congestion and ramping surpluses.  It is also shown that, comparing with LMP,  TLMP imposes a penalty on generators with limited ramping capabilities, thus giving incentives for generators to reveal their ramping limits truthfully and improve their ramping capacities.  Several benchmark pricing mechanisms are evaluated under the rolling-window dispatch and pricing models. The performance measures considered are the level of out-of-the-market uplifts, the revenue adequacy of the system operator, consumer payment, generator profit, level of discriminative payment, and price volatility. 
\end{abstract}

\begin{IEEEkeywords}
Multi-interval economic dispatch, look-ahead dispatch,  locational marginal pricing, general and partial equilibrium, and dispatch-following incentives.
\end{IEEEkeywords}

\section{Introduction} \label{sec:intro}
This two-part paper addresses some of the open problems in pricing multi-interval dispatch subject to ramping constraints and forecasting uncertainty.  Part I focuses on theoretical issues surrounding dispatch-following incentives with three major conclusions.  One is that, under the rolling-window dispatch model, uniform-pricing schemes cannot provide dispatch-following incentives that avoid out-of-the-market uplifts. Because such uplifts are discriminative, price-discrimination is necessary.

Another conclusion is that the temporal locational marginal pricing (TLMP)---a generalization of locational marginal pricing (LMP)---provides full dispatch-following incentives that eliminate the need for out-of-the-market uplifts under the rolling-window economic dispatch model and arbitrary demand forecast accuracy.

The third conclusion is that it is optimal for price-taking profit maximizing generators to bid with their true marginal costs of generation.   

Providing dispatch-following incentive is only one of the many measures that pricing mechanisms need to be evaluated for adoption. This paper presents a study on a broader range of issues relating to pricing multi-interval dispatch under a more general network model.

We focus on two categories of performance measures.  The first is on incentive compatibilities specific to multi-interval dispatch.   One type of incentive is the degree for which a particular pricing mechanism provides the necessary dispatch-following incentives for generators.  Here we measure the lack of dispatch-following incentives by the size of the (ex-post) lost-of-opportunity (LOC) payment.  The higher the LOC payment,  the greater the incentive for a generator to deviate from the dispatch signal. The other type is the incentive for a generator to reveal its ramping limits truthfully.  If a generator receives higher profit under a pricing mechanism by under-reporting ramping capability, then the pricing mechanism not only distorts the actual ramping ability of the system but also discourages a generator from improving its ramping capacity.

The second category of performance measures is on the revenue adequacy and social welfare distribution.  We are particularly interested in whether a pricing mechanism ensures the revenue adequacy of the independent system operator (ISO).  For multi-interval dispatch over a network with power flow constraints,  the revenue of the operator needs to cover the generation cost,  the cost of ramping-induced out-of-the-market uplifts, and the congestion rent.   Since the operator is regulated to be revenue-neutral, a revenue reconciliation process typically redistributes the surplus or shortfall of the system operator to its consumers.
Thus the revenue adequacy of the operator affects costs to consumers.

Different pricing mechanisms result in different allocations of social welfare.   A pricing scheme that yields higher generator profits may be more costly to consumers.  In general, no pricing mechanism dominates all others across a wide range of performance measures.  A regulator of public utility typically favors a pricing policy that guarantees the generators' revenue adequacy while minimizing the cost to consumers.

Transparency and volatility are also relevant metrics for evaluating pricing mechanisms.  Uniform pricing schemes are transparent and effective pricing signals for market participants.  The use of out-of-the-market uplifts, however, affects the transparency of uniform pricing.  Nonuniform pricing, in general, lacks transparency.

\subsection{Summary of results and related work}
The main contribution of Part II is twofold. First, we extend key theoretical results in Part I to a network setting in Proposition~\ref{prop1:RevenueLMP}-\ref{prop5:EQTLMP}.
Whereas most theoretical results such as the strong equilibrium
property of TLMP generalize naturally to systems with
network constraints, we obtain new results that demonstrate
succinctly the spatial-temporal decomposition of TLMP.

Proposition~\ref{prop2:TLMP} gives an explicit decomposition of TLMP into energy, congestion, and ramping prices, which shows that TLMP is the sum of a public price in the form of {locational marginal price (LMP)} and a private ramping price; the former is transparent to all participants, and the latter plays the role of in-the-market discrimination among generators with different ramping capabilities.

We show in Proposition~\ref{prop3:RevenueTLMP} that, under the one-shot economic dispatch model with perfect demand forecast, the merchandising surplus under TLMP is positive and is equal to the sum of the congestion surplus (congestion rent) and the ramping surplus defined by the surplus due to binding ramping constraints.  In contrast, Proposition~\ref{prop1:RevenueLMP}  shows that the merchandising surplus of LMP covers only the congestion rent.  This result explains partially that the revenue of the operator under LMP is often inadequate to cover the out-of-the-market uplifts due to binding ramping constraints.

Proposition~\ref{prop3:RevenueTLMP} also shows that the payment to a generator under LMP is higher than that under TLMP. And the price decomposition of TLMP in (\ref{eq:TLMP2}) implies that the ramping price of TLMP imposes penalty on generators for their inabilities to support the system's ramping needs. 
From an individual generator's perspective, Proposition~\ref{prop4:RevenueGap} shows that, under TLMP,  a generator with higher ramping limit receives higher payment than an identical generator with limited ramping capability.   This result partially explains that TLMP discourages under-reporting ramping limits and encourages generators to improve ramping capabilities.

The second part of our contribution is the empirical simulation studies on incentives, the revenue adequacy of the ISO, consumer payments, generator profits, and the level of discriminative payments. We are particularly interested in the effects of forecasting errors and congestions
on these performance measures.  We compared several benchmark pricing schemes in the literature under the rolling-window dispatch model: the classical multi-interval LMP, TLMP, price-preserving multi-interval pricing (PMP) \cite{Hogan:16,Hua&etal:19TPS}, constraints-preserving multi-interval pricing (CMP) \cite{Hua&etal:19TPS}, and multi-settlement LMP (MLMP)  \cite{Zhao&Zheng&Litvinov:19arxiv}.

There is a fairly extensive literature on pricing multi-period dispatch.  See a summary of related work in Part I \cite{Guo&Chen&Tong:19arxiv} and references therein.  The impact of multi-interval dispatch on LMP was considered in \cite{Thatte&Choi&Xie:14PSCC}.
The works most relevant to this paper are recent works of Hua \etal \cite{Hua&etal:19TPS} and Zhao, Zheng, and Litvinov \cite{Zhao&Zheng&Litvinov:19arxiv} that articulate some of the critical issues and set forth formal statements of investigation.

Proofs,  some detailed derivations, and additional simulations involving network constraints can be found in the appendix at the end of this paper.

\section{System and operation models} \label{sec:model}
\subsection{Generation, demand, and network models}
We consider a power system with $M$ buses under the direct-current (DC) power flow model with line-flow constraints.  We follow the same notations used in Part I, adding bus indices as superscripts to relevant variables.

Without loss of generality, we assume that every bus has  $N$ generators\footnote{One use non-generating generators to make up total $N$ generators by setting the generation capacites to zero of such generators.}.  Let $g_{it}^{m}$ be the dispatch of generator $i$ at bus $m$ in interval $t$, $\gbf^m[t]=(g_{1t}^m,\cdots,g_{Nt}^m)$ the dispatch vector at bus $m$, and $\gbf[t]=(\gbf^1[t],\cdots,\gbf^M[t])$ the dispatch vector in interval $t$  from all generators.

We assume that there is one aggregated inelastic demand at each bus.  For the demand at bus $m$, let $d_t^m$ be the actual demand in interval $t$,  $\hat{d}_t^m$ the forecasted demand, $\dbf[t]=(d_t^1,\cdots, d_t^M)$ the demand vector from all buses in interval $t$, and $\hat{\dbf}[t]$ the forecast of $\dbf[t]$.

The spatial property of the  power flow is governed by the DC power flow model where the branch power flow vector is a linear function of the {\em net power injection} $(\qbf[t]-\dbf[t])$ where $\qbf[t]=(q_t^1,\cdots,q_t^M)$ is the vector of bus generations, and $q_t^m=\sum_i g_{it}^m$  the total generation from bus $m$ in $t$.

For a network with total $B$ branches,  the $2B$-dimensional vector $\zbf[t]$ of branch power flows\footnote{Each branch has two directional power flows.} satisfies 
\[
\zbf[t] = \Sbf (\qbf[t]-\dbf[t]),
\] 
where $\Sbf$ is the $2B\times (M-1)$ shift-factor matrix\footnote{Matrix $\Sbf$ can be made time varying without affecting the results. Note that a slack bus should be removed in matrix S.}.


 \subsection{The rolling-window dispatch model}
 The rolling-window economic dispatch (R-ED) policy $\Gc^{\RED}$  is defined by a sequence of $W$-interval look-ahead economic dispatch policies $(\Gc_t^{\RED}, t=1, \cdots, T)$.

 At time $t$,  $\Gc^{\RED}_t$ solves the following $W$-interval economic dispatch optimization using  (i) the realized dispatch $\gbf^{\RED}[t-1]$ in interval $t-1$ and (ii) the load forecast $(\hat{\dbf}[t],\cdots,\hat{\dbf}[t+W-1])$ in $W$ intervals, assuming that the forecast in the binding interval $t$ is perfect, \ie $\hat{\dbf}[t]=\dbf[t]$.
\beq \label{eq:RED}
\begin{array}{lrl}
 & \Gc^{\RED}_t: & \mbox{at time $t$,}  \\[0.2em]
 &\underset{\{\Gbf=[g_{it}^m]\}}{\rm minimize}   &  F_t(\Gbf)  \\
&  \mbox{subject to:} & \mbox{{ Network constraints:}}\\
& \lambda_{t'}: & \sum_{m=1}^M \sum_{i=1}^N g_{it'}^m = \sum_{m=1}^M \hat{d}_{t'}^m,\\
&\pmb{\phi}[t']: & \Sbf (\qbf[t']-\hat{\dbf}[t']) \le \cbf,\\
 & & \hfill \mbox{for all $t\le t'\ < t+W.$}\\[0.2em]
 &  & \mbox{ Generation constraints:}\\
& (\underline{\mu}^m_{it'},\bar{\mu}^m_{it'}):  &  -\underline{r}^m_i\le g^m_{i(t'+1)}-g^m_{it'} \le \bar{r}^m_{i},\\
& (\underline{\rho}^m_{it'},\bar{\rho}^m_{it'}):   & 0 \le g^m_{it'} \le \bar{g}^m_{i}, \\
 & & \hfill \mbox{for all  $m, t\le t' < t+W$.}\\
  &  & \mbox{Boundary ramping constraints:}\\
  & \bar{\mu}^m_{i(t-1)}:  &  g^m_{it}-g^{\RED}_{mi(t-1)} \le \bar{r}^m_{i},\\
  &\underline{\mu}^m_{i(t-1)}: & g^{\RED}_{mi(t-1)} - g^m_{it} \le  \underline{r}^m_i,\\
   & & \hfill \mbox{for all  $m$.}\\
\end{array}
\eeq
where $\Gbf=[\gbf[t],\cdots, \gbf[t+W-1]] $ is the matrix of all generation variables in the $W$-interval look-ahead window,  and $F_t(\Gbf)$ is the total bid-in costs
 \[
 F_t(\Gbf) :=\sum_{i=1}^N\sum_{m=1}^M\sum_{t'=t}^{t+W-1} f_{it'}^m (g_{it'}^m).  
 \]
 Here $f_{it'}^m(\cdot)$ is the bid-in cost of  generator $i$ at bus $m$ in interval $t'$, assumed to be convex and piecewise linear (or quadratic). Vector $\cbf\ge {\bf 0}$ is the vector of line-flow limits.

Dual variables in (\ref{eq:RED}) play a prominent role in multi-interval pricing, where $\lambda_{t'}$  is the dual variable associated with the power balance equation in interval $t'$, $\pmb{\phi}[t']$  the dual variables associated with line constraints, and $(\underline{\mu}^m_{it'},\bar{\mu}^m_{it'},\underline{\rho}^m_{it'},\bar{\rho}^m_{it'})$ the dual variables for the lower and upper limits for ramping and generation, respectively.

Let $(g_{it'}^{m*})$ be the solution to the above optimization, and $\underline{\mu}^{m*}_{i(t-1)}, \bar{\mu}^{m*}_{i(t-1)}$ the optimal dual variables.
Under R-ED policy $\Gc_t^{\RED}$, the dispatch in the binding  interval $t$ is set at
\beq \label{eq:gED}
g_{mit}^{\RED}:= g_{it}^{m*}.
\eeq
Also relevant are the (shadow) ramping prices $(\underline{\mu}^{m*}_{i(t-1)},\bar{\mu}^{m*}_{i(t-1)})$
 that capture the interdependencies of decisions across sliding windows.
 For later references, define the boundary ramping prices as
 \beq
\underline{\mu}^{\RED}_{mit}:=\underline{\mu}^{m*}_{i(t-1)}, ~~\bar{\mu}^{\RED}_{mit}:=\bar{\mu}^{m*}_{i(t-1)}.
\eeq

In contrast to the rolling-window dispatch, the  {\em one-shot economic dispatch} $\Gc^{\mbox{\tiny 1-ED}}$  produces the dispatch of the entire scheduling period at once using the solution $\Gbf^*$ of  (\ref{eq:RED}) at $t=1$ and window size $W=T$.

\section{Rolling-window LMP and TLMP} \label{sec:TLMP}
A rolling-window pricing policy $\Pc=(\Pc_1,\cdots, \Pc_T)$ follows the same structure as the rolling-window economic dispatch.  At time $t$, $\Pc_t$    sets  prices at all $M$ buses for the binding interval $t$.  It may also provide advisory prices for the future intervals within the pricing window $\Hmsc_t=\{t,\cdots, t+W-1\}$.

 Here we generalize the standard rolling-window LMP (R-LMP) policy $\Pc_t^{\RLMP}$  and the rolling-window TLMP $\Pc_t^{\RTLMP}$ derived in Part I for systems with power flow constraints. Both R-LMP and R-TLMP are marginal cost pricing mechanisms derived from the R-ED optimization (\ref{eq:RED});  they are by-products of the R-ED policy.

\subsection{Rolling-window LMP (R-LMP) and Properties}
Let the realized price vector  in the binding interval $t$ set by  R-LMP  be $\pibf^{\RLMP}[t]=(\pi^{\RLMP}_{1t},\cdots,\pi^{\RLMP}_{Mt})$ where $\pi_{mt}^{\RLMP}$ is the uniform price  for all generators and demand at bus $m$.

The R-LMP $\pi_{mt}^{\RLMP}$  is defined by the marginal cost of meeting demand $d^m_t$ at bus $m$ in interval $t$.  From (\ref{eq:RED}) and by the envelope theorem, we have
\beq \label{eq:RLMP}
\pibf^{\RLMP} [t] = \nabla_{\hat{\dbf}[t]} F_t(\Gbf) = \lambda^{\RLMP}_t {\bf 1} - \Sbf^{\intercal} \pmb{\phi}^{\RLMP}[t],
\eeq
where  ${\bf 1}$ is a vector of $1$'s, $\lambda^{\RLMP}_t$ and $\pmb{\phi}^{\RLMP}[t]$ the shadow prices\footnote{When defining prices with Lagrange multipliers, we implicitly assume that the solutions to the dual optimization are unique.} from (\ref{eq:RED}) for the power balance and  congestion constraints in interval $t$, respectively.

We summarize next main properties of R-LMP.  Even though R-LMP is computed based on the current and future demand forecasts subject to ramping constraints, many properties of the single-period LMP hold for the multi-interval R-LMP.

\subsubsection{Energy-congestion price decomposition}
The R-LMP expression (\ref{eq:RLMP})  shows an explicit energy-congestion price decomposition, where the first term $\lambda_t^{\RLMP}$ is the system-wide uniform-price of energy for all generators and demands.  The second term $\Sbf^{\intercal} \pmb{\phi}^{\RLMP}[t]$ is the congestion-induced price discrimination at different locations. Note that there are no ramping prices explicitly shown in R-LMP; the R-LMP expression is identical to that in the standard single-interval LMP.  The inter-temporal effects of ramping on R-LMP are hidden in the sequence of R-LMP prices $\pibf^{\RLMP}[t]$.

\subsubsection{Equilibrium properties}
We have shown in Part I that, for the single-bus network and under the perfect load forecast assumption, the one-shot economic dispatch $\Gbf^{\ED}$ and LMP $\pibf^{\LMP}$ form a general equilibrium.  This property holds for systems with network constraints.  Unfortunately, the rolling-window version of economic dispatch and LMP $(\gbf^{\RED},\pibf^{\RLMP})$ do not satisfy the general equilibrium condition in general, even when the load forecasts are accurate;
out-of-the-market uplifts are necessary.

\subsubsection{ISO's revenue adequacy}  The classical LMP theory for the single-interval LMP policy \cite{Wu&Variaya&Spiller&Oren:96JRE} states that the ISO has a non-negative merchandising surplus that covers and  {\em only covers} the system congestion rent.   This result extends to R-LMP  under arbitrary forecast errors when there are ramping constraints. 

\begin{proposition}[ISO revenue adequacy under R-LMP] \label{prop1:RevenueLMP}
For all  $(\gbf^{\RED}[t],\pibf^{\RLMP}[t])$ generated by the  R-ED and R-LMP policies  under arbitrary forecasting errors, the ISO  has non-negative merchandising surplus
\bea
\mbox{\rm MS}^{\RLMP} = \sum_{t=1}^T \cbf^{\intercal} \pmb{\phi}^{\RLMP}[t] \ge 0.\nn
\eea
\end{proposition}
Proposition~\ref{prop1:RevenueLMP} shows that the merchandising surplus from R-LMP covers and {\em only covers} the congestion rent designated to pay transmission-line owners and financial transmission right (FTR) holders.   There is no extra surplus within the market settlement to cover the out-of-the-market uplifts designed to ensure dispatch-following incentives. Thus the ISO is likely to be revenue inadequate under R-LMP  when the ISO has to pay out-of-the-market uplifts.

\subsection{Rolling-window TLMP (R-TLMP) and Properties}
As a generalization of R-LMP to a nonuniform marginal-cost pricing, R-TLMP  allows individualized prices for generators and demands.   Specifically, the R-TLMP at bus $m$ in interval $t$ is a set of prices
\[
\pibf_{m}^{\RTLMP}[t]=(\pi^{\RTLMP}_{m0t}, \pi^{\RTLMP}_{m1t},\cdots, \pi^{\RTLMP}_{mNt}),
\]
where $\pi^{\RTLMP}_{m0t}$ is the price for the demand and $\pi^{\RTLMP}_{mit}$ the price for generator $i$ at bus $m$.

For the demand at bus $m$ in interval $t$, its R-TLMP $\pi_{m0t}^{\RTLMP}$  is defined as the marginal cost to the system to satisfy the demand $d^m_t$---the same definition used in LMP:
\[
\pi^{\RTLMP}_{m0t} := \frac{\partial}{\partial \hat{d}_t^m} F_t(\Gbf) = \pi_{mt}^{\RLMP}.
\]
The R-TLMP for  generator $i$ at bus $m$, on the other hand, is defined by the {\em marginal benefit} of generator producing power $g_{it}^{m*}$. In other words, generator $i$  is treated as an inelastic negative-demand set at the R-ED solution to (\ref{eq:RED}), \ie $g_{it}^m=g_{it}^{m*}$. As defined in Part I,
\[
\pi^{\RTLMP}_{mit} := - \frac{\partial}{\partial g_{it}^m} F_{-it}^m(\Gbf^*),
\]
where $F_{-it}^m(\Gbf)=F_t(\Gbf)-f_{it}^m(g_{it}^m)$ is the total generation cost excluding that from generator $i$ at bus $m$  in interval $t$.  

The following proposition generalizes the TLMP  expression in Part I.

\begin{proposition}[Price decomposition of R-TLMP]  \label{prop2:TLMP}
Let  $(\lambda^*_t,\pmb{\phi}^*[t], \underline{\mu}_{i(t-1)}^{m*}, \bar{\mu}_{i(t-1)}^{m*}, \underline{\mu}_{it}^{m*}, \bar{\mu}_{it}^{m*})$ be the optimal values of the dual variables associated with  the constraints in (\ref{eq:RED}).

The R-TLMP for the demand $\hat{d}^m_t$ at bus $m$ in interval $t$ is given by
 \beq
\pi^{\RTLMP}_{m0t} =  \lambda^*_{t} -\sbf_m^{\intercal}\pmb{\phi}^*[t]=\pi_{mt}^{\RLMP},
\eeq
where
where $\sbf_m$ is the $m$-th column of the shift-factor matrix $\Sbf$ corresponding to bus $m$.

The R-TLMP for generator $i$ at bus $m$ in interval $t$ is given by
\bea \label{eq:RTLMP}
\pi^{\RTLMP}_{mit}  &=& \lambda^*_t -\sbf_m^{\intercal}\pmb{\phi}^*[t] + \Delta_{it}^{m*} \label{eq:TLMP1}\\
&=& \pi_{mt}^{\RLMP} + \Delta_{it}^{m*}, \label{eq:TLMP2}
\eea
where $\Delta_{it}^{m*} = \Delta \mu_{it}^{m*}-\Delta \mu_{i(t-1)}^{m*} $, and $\Delta \mu_{it}^{m*}:=\bar{\mu}_{it}^{m*}-\underline{\mu}_{it}^{m*}$.
\end{proposition}

Properties of R-TLMP for power systems with network constraints are summarized next.

\subsubsection{Energy-congestion-ramping decomposition}
The specific form of R-TLMP in (\ref{eq:TLMP1}) reveals an explicit space-time decomposition of payment to generators:  a system-wide uniform energy price in $\lambda^*_t$ applies to all generators and demands everywhere, a spatial discriminative price in the form of location-specific congestion prices in $\sbf_m^{\intercal}\pmb{\phi}^*[t]$ applying to all generators and demands at bus $m$, and a generator-specific temporal ramping prices in $\Delta_{it}^{m*}$ that serves as a ``penalty'' to the generator for its limited ramping capability.   The penalty interpretation of $\Delta_{it}^{m*}$ is especially important for the incentives of the truthful revelation of ramping limits, as discussed next.

\subsubsection{Public-private price decomposition and transparency}
The structure of R-TLMP shown in (\ref{eq:TLMP2})  shows a public and private price decomposition: the R-LMP part of R-TLMP captures the standard uniform pricing for the energy and congestion costs that are transparent to all market participants. By revealing the R-LMP part of the TLMP, the system operator can provide the necessary system-wide pricing signal effectively for market participants.

On the other hand, the ramping price $\Delta_{it}^{m*}$ of R-TLMP is private; it pertains to  the ramping conditions of individual generators. It is neither necessary nor practical to make this part of the price transparent.  Another interpretation of $\Delta_{it}^{m*}$  is that it plays the role of uplift payments for uniform prices that ensures dispatch following incentives for the generator, except that it is computed within the real-time market.  It is in this interpretation that R-TLMP has the same level of transparency of all uniform pricing schemes that require out-of-the-market uplifts.

\subsubsection{ISO's revenue adequacy} The space-time decomposition of R-TLMP provides insights into sources of ISO's surplus.  To this end, we consider the ideal case of one-shot TLMP with a perfect load forecast.

\begin{proposition}[ISO revenue adequacy under TLMP] \label{prop3:RevenueTLMP}
Consider the one-shot economic dispatch $\Gc^{\oneED}$  defined in (\ref{eq:RED}) with $t=1$, $W=T$  and perfect demand forecast.  Let   the solution of the dual variables associated with  the constraints be $(\lambda^*_t,\pmb{\phi}^*[t],  \underline{\mu}_{it}^{m*}, \bar{\mu}_{it}^{m*})$.
The total ISO merchandising surplus decomposes into  ramping  and  congestion surpluses:
\bea
\mbox{\rm MS}^{\TLMP} &=& \mbox{\rm MS}^{\mbox{\tiny\rm ramp}} + \mbox{\rm MS}^{\mbox{\tiny\rm con}},\label{eq:MS_TLMP}
\eea
where
\bea
 \mbox{\rm MS}^{\mbox{\tiny\rm ramp}} &=&\sum_{m,i,t} ( \bar{\mu}^{m*}_{it}\bar{r}^m_{i} + \underline{\mu}^{m*}_{it}\underline{r}^m_{i}) \ge 0,\\
 \mbox{\rm MS}^{\mbox{\tiny\rm con}} &=& \sum_{t} \cbf^{\intercal}\pmb{\phi}^*[t] \ge 0.
\eea
\end{proposition}
The above proposition does not generalize to the rolling-window TLMP policy, unfortunately. There are indeed cases when TLMP does not guarantee revenue adequacy (after the congestion surplus is removed).  Nonetheless, simulations show that the shortfall in TLMP is considerably smaller than those of its alternatives.

\subsubsection{Ramping price as a penalty for inadequate ramping}
Note that the TLMP and  LMP have the same demand price (thus the same revenue) and the same congestion surplus.  From (\ref{eq:MS_TLMP}) and the fact that $\mbox{\rm MS}^{\LMP} = \mbox{\rm MS}^{\mbox{\tiny\rm con}}$,  the total generator payment under TLMP  must be less than that under LMP.     The following proposition suggests that the ramping price $\Delta_{it}^{m*}$ of TLMP plays the role of penalty for inadequate ramping.

\begin{proposition}[Revenue gap under LMP and TLMP] \label{prop4:RevenueGap}
Consider the one-shot economic dispatch $\Gc^{\oneED}$  defined in (\ref{eq:RED}), and let $(g_{it}^{m*}, \underline{\mu}_{it}^{m*}, \bar{\mu}_{it}^{m*})$ be the solution of the primal and dual variables associated with  generator $i$ at bus $m$  and interval $t$.  If\footnote{The assumption $\mu_{i0}^{m*}=\mu_{iT}^{m*}=0$ has minimum impact for large $T$ and becomes innocuous if the initial and final ramping constraints can be relaxed.}  $\mu_{i0}^{m*}=\mu_{iT}^{m*}=0$,  then the revenue difference for delivering $(g_{it}^{m*}, t=1,\cdots, T)$ under LMP and TLMP is nonnegative and
\bea
R_{mi}^{\LMP} - R_{mi}^{\TLMP}
&=& \bar{r}_i^m \sum_t \bar{\mu}_{it}^{m*} + \underline{r}_i^m\sum_t \underline{\mu}_{it}^{m*} \ge 0. \nn
\eea
\end{proposition}
For an interpretation, consider two generators at the same bus with the same generation level. One generator has high ramping limits so that there are no binding ramping constraints; the other has binding ramping constraints.  Under LMP, the two generators receive the same payment.  Proposition~\ref{prop4:RevenueGap} shows that, under TLMP, however,  the one with high ramping limits receives a higher payment than the one having binding ramping constraints.  This suggests that it is to the generator's benefit not to under-report its ramping limit, and the generator is incentivized to improve its ramping capability. This insight is validated in simulations in Sec~\ref{sec:Perf}.

\subsubsection{Equilibrium properties}
The strong equilibrium property of R-TLMP shown in Part I holds when network constraints are imposed. Under TLMP, there is no incentive for any generator to deviate from the dispatch signal regardless of the accuracy of demand forecast and no need for out-of-the-market uplifts.
\begin{proposition}[Strong equilibrium property of TLMP]\label{prop5:EQTLMP}
For every  load forecast, let $\Gbf^{\RED}$ and $\pibf^{\RTLMP}$ be the rolling-window economic dispatch and the rolling-window TLMP, respectively.  Then $(\Gbf^{\RED},\pibf^{\RTLMP})$ satisfies the strong equilibrium conditions that result in  zero LOC uplifts.
\end{proposition}

\section{Related Benchmark Pricing Policies} \label{sec:Benchmark}
We present here several benchmark pricing policies that also use the same rolling-window dispatch model.  Missing in the discussion is the flexible ramping product (FRP) that has been implemented in CAISO because FRP uses a different optimization procedure that produces different dispatch signals.
The development here follows \cite{Hua&etal:19TPS,Schiro:17,Zhao&Zheng&Litvinov:19arxiv}.

\subsection{Price-Preserving Multi-interval Pricing (PMP)}
Unlike LMP and TLMP that derive prices from R-ED, PMP \cite{Hogan:16,Hua&etal:19TPS} employs a separate pricing optimization aimed at minimizing the uplift payment.

The rolling-window PMP policy $\Gc^{\RPMP}_t$ at time $t$  sets uniform prices $\pibf^{\RPMP}[t]$ in the binding interval $t$ using (i) the past rolling-window PMP prices\footnote{In practical implementation, one may include only a few past decision intervals.} $(\pibf^{\RPMP}[t-1],\cdots,\pibf^{\RPMP}[1])$  and (ii)  the demand forecasts $(\hat{\dbf}[t],\cdots, \hat{\dbf}[t+W-1])$ in the look-ahead window.

At time $t$, let $\Gbf=\big[\gbf[1],\cdots, \gbf[t+W-1]\big]$ be all the generation variables involved in the past, current, and look-ahead intervals.  The rolling-window PMP policy $\Gc^{\RPMP}_t$ solves the following optimization:
\beq \label{eq:PMP}
\begin{array}{lrl}
 & \Gc^{\RPMP}_t:& \mbox{at time $t$,} \\[0.2em]
&\underset{\Gbf \in \Gmsc^{\RPMP}}{\rm minimize}   &  F(\Gbf) -\sum_{t'=1}^{t-1} \qbf^{\intercal}[t']\pibf^{\RPMP}[t']\\
&  \mbox{subject to:} & \mbox{for all $t\le t' < t+W$}\\
& & \qbf[t']=(\sum_i g_{it'}^1,\cdots,\sum_i g_{it'}^M),\\
& \lambda_{t'}: & {\bf 1}^{\intercal} \qbf[t'] ={\bf 1}^{\intercal} \hat{\dbf}[t']\\
&\pmb{\phi}[t']: & \Sbf (\qbf[t']-\hat{\dbf}[t']) \le \cbf,
\end{array}
\eeq
where  $\Gmsc^{\RPMP}$ represents the set of  individual generation constraints such as ramp and generation limits.  See Appendix F of this paper.

The rolling-window PMP sets the price for generation in interval $t$ by
\beq \label{eq:PMP1}
\pibf^{\RPMP}[t] = \lambda^{\RPMP}_t {\bf 1}- \Sbf^{\intercal}\pmb{\phi}^{\RPMP}[t],
\eeq
where $\lambda^{\RPMP}_t$ and $\pmb{\phi}^{\RPMP}[t]$ are the multipliers
associated with power balance and line-flow constraints in (\ref{eq:PMP}).

Note that the objective function can be written as
\[
\sum_{t'=1}^{t+W-t}\sum_{m,i} f_{it'}^m (g_{it'}^m) -\sum_{t'=1}^{t-1}\sum_{m,i} (\pi_{mit'}^{\RPMP} g_{it'}^m-f_{it'}^m(g_{it'}))
\]
where the first term is the (bid-in) generation cost in the look-ahead window.  Ignoring the first term, the second term (without the negative sign) represent the estimate of the total surplus (including the LOC uplifts) up to time $t-1$.

%
%


\subsection{Constraint-Preserving Multi-interval Pricing (CMP)}
CMP \cite{Hua&etal:19TPS} is another policy that generates uniform prices in a separate optimization different from the rolling-window economic dispatch.  Instead of involving past settled prices in PMP, CMP enforces the ramping constraints between the rolling-window economic dispatch and the dispatch variables used in the pricing models.

The rolling-window CMP policy $\Gc^{\RCMP}_t$ at time $t$  sets prices $\pibf^{\RCMP}[t]$ in the binding interval $t$ using (i) the past rolling-window economic dispatch $\gbf^{\RED}[t-1]$, (ii) shadow prices from (\ref{eq:RED})  $(\underline{\mu}_{mit}^{\RED},\bar{\mu}_{mit}^{\RED})$  that tie generation between intervals $t-1$ and $t$, and (iii) load forecasts $(\hat{\dbf}[t],\cdots, \hat{\dbf}[t+W-1])$ in the look-ahead window.

Let $\Gbf=\big[\gbf[t],\cdots, \gbf[t+W-1]\big]$ be the generation variables within the $W$-interval lookahead window, and $F_t(\Gbf)$ the total cost of generation.
 The rolling-window CMP policy $\Gc_t^{\RCMP}$ solves the following optimization:
\beq \label{eq:RCMP}
\begin{array}{lrl}
 & \Gc^{\RCMP}_t:& \mbox{at time $t$,}\\[0.2em]
 &\underset{\Gbf \in \Gmsc^{\RCMP}}{\rm minimize}   &  F_t(\Gbf)  + \sum_{m,i} (\bar{\mu}_{mit}^{\RED}-\underline{\mu}_{mit}^{\RED}) g^m_{it}\\
 &  \mbox{subject to:} & \mbox{for all $t\le t' < t+W$}\\
 & & \qbf[t']=(\sum_i g_{it'}^1,\cdots,\sum_i g_{it'}^M),\\
& \lambda_{t'}: & {\bf 1}^{\intercal} \qbf[t'] ={\bf 1}^{\intercal} \hat{\dbf}[t']\\
&\pmb{\phi}[t']:  & \Sbf (\qbf[t']-\hat{\dbf}[t']) \le \cbf,\\
\end{array}
\eeq
where  $\Gmsc^{\RCMP}$ represents the set of individual generation constraints.  See Appendix F of  of this paper.

Let $\lambda_t^{\RCMP}, \pmb{\phi}^{\RCMP}[t]$ be the dual variable solution to the above optimization associated with the power balance equation and line flow constraints, respectively.
The rolling-window CMP set the price at bus $m$ and interval $t$  by
\beq \label{eq:RCMP1}
\pibf^{\RCMP}[t]=\lambda_t^{\RCMP}{\bf 1} - \Sbf^{\intercal}\pmb{\phi}^{\RCMP}[t].
\eeq

\subsection{Multi-settlement LMP (MLMP)}
The multi-settlement LMP extends the two-settlement LMP used in the day-ahead and real-time markets to the rolling-window dispatch setting.  
\vspace{-2em}
\begin{figure}[h]
\center
\begin{psfrags}
\psfrag{G1}[l]{\normalsize Settlement 1}
\psfrag{H1}[l]{\small $\Hmsc_{t^*-2}$}
\psfrag{G2}[l]{\normalsize  Settlement 2}
\psfrag{H2}[l]{\small $\Hmsc_{t^*-1}$}
\psfrag{G3}[r]{\normalsize Settlement $W$}
\psfrag{H3}[r]{\small $\Hmsc_{t^*}$}
\psfrag{t0}[l]{\small $t^*$}
\psfrag{t1}[l]{\small $t^*-1$}
\psfrag{t2}[l]{\small $t^*-2$}
\psfrag{a}[c]{\small $\begin{array}{c}\hat{g}_{it^*}^{m,1}\\\hat{\pi}_{t^*}^{m,1}\end{array}$}
\psfrag{b}[c]{\small $\begin{array}{c}\hat{g}_{it}^{m,2}\\\hat{\pi}_{t^*}^{m,2}\end{array}$}
\psfrag{c}[c]{\small $\begin{array}{c}\hat{g}_{it^*}^{m,W}\\\hat{\pi}_{t^*}^{m,W}\end{array}$}
\scalefig{0.38}\epsfbox{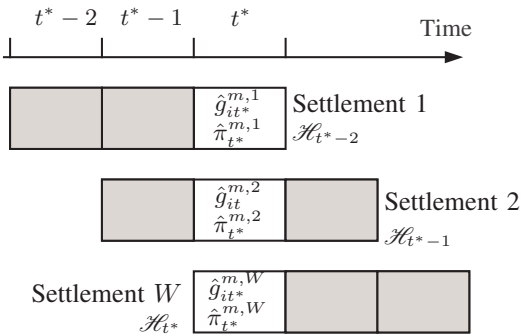}
\end{psfrags}
\caption{\small Rolling-window dispatch with window size $W=3$.  The final generation and payments are determined in $W=3$ settlements, each produces the generation quantities and prices for deviation from the quantity in the previous settlement.}
\label{fig:MS}
\end{figure}

We use Fig.~\ref{fig:MS} to illustrate the settlement process for generation and demand in interval $t^*$.  When pricing the $W$-interval rolling-window dispatch $g_{mit^*}^{\RED}$ for generator $i$ at bus $m$, we consider $W$  settlements from $W$  sequential ``markets'', one for each sliding window that includes interval $t^*$ as shown in Fig.~\ref{fig:MS}.

The first settlement occurs at time $t=t^*-W+1$ with scheduling window $\Hmsc_{t^*-W+1} = \{t^*-W+1, \cdots, t^*\}$.    Let  $\hat{g}_{it^*}^{m,1}$  be the advisory dispatch for generator $i$ at bus $m$ in interval $t^*$ computed by the $W$-interval economic dispatch (\ref{eq:RED}) and $\hat{\pi}_{t^*}^{m,1}$  its LMP.  Here the superscript ``1'' indicates that this is the first market  that the dispatch in interval $t^*$ is settled
 financially.  The first financially binding settlement for generator $i$ at bus $m$  is  $\hat{\pi}_{t^*}^{m,1} \times \hat{g}_{it^*}^{m,1}$ (\$) for the advisory dispatch $\hat{g}_{it^*}^{m,1}$  in interval $t^* $.  (This settlement is analogous to the day-ahead settlement in the two-settlement process.)

The second settlement for generator $i$ at bus $m$  occurs at time $t^*-W+2$ using the rolling-window dispatch over scheduling window $\Hmsc_{t^*-W+2}$.  Let
$(\hat{g}_{it^*}^{m,2}, \hat{\pi}_{it^*}^{m,2})$ be the dispatch-LMP pair computed by the  economic dispatch over $\Hmsc_{t^*-W+2}$.  The second financially binding settlement for generator $i$ at bus $m$ is  $\hat{\pi}_{t^*}^{m,2} \times (\hat{g}_{it^*}^{m,2}-\hat{g}_{it^*}^{m,1})$ (\$) for the advisory dispatch of $\hat{g}_{it^*}^{m,2}$ in interval $t^*$.

As the window slides forward one interval at a time,  the process  generates a sequence of $W$ dispatch-LMP pairs $(\hat{g}_{it^*}^{m,1},\hat{\pi}_{t^*}^{m,1}),  \cdots, (\hat{g}_{it^*}^{m,W},\hat{\pi}_{t^*}^{m,W})$  for generator $i$ at bus $m$.    In the last settlement occurs  at  time  $t=t^*$  when 
 generator $i$ at bus $m$ physically delivers $\hat{g}_{it^*}^{m,W}=g_{mit^*}^{\RED}$ and receives the final settlement $\hat{\pi}_{t^*}^{m,W}\times(\hat{g}_{it^*}^{m,W}-\hat{g}_{it^*}^{m,W-1})$ (\$).  Note that  $\hat{\pi}_{t^*}^{m,W}=\pi_{m,t}^{\RLMP}$.

Under the multi-settlement LMP, the total  revenue  $R^{\MLMP}_{mit^*}$  for generator $i$ at bus $m$   for delivering power $g_{mit^*}^{\RED}$ is 
\beq
R^{\MLMP}_{mit^*} = \hat{\pi}_{t^*}^{m,1}(\hat{g}_{it^*}^{m,1}) +  \sum_{k=2}^W \hat{\pi}_{t^*}^{m,k}(\hat{g}_{it^*}^{m,k}- \hat{g}_{it^*}^{m,k-1}).
\eeq
Note that, although $R^{\MLMP}_{mit^*} $ is a linear function with respect to $(\hat{g}_{it^*}^{m,1},\cdots, \hat{g}_{it^*}^{m,W})$, it is not  linear with respect to the  power delivered $g_{mit^*}^{\RED}$ in interval $t^*$.

\section{Performance} \label{sec:Perf}
We present here  simulation results involving three generators at a single bus.   Simulations for larger networks including one involving an ISO-NE 8-zone case can be found in the appendix of this paper. In all our simulations, we have quantities in MW and prices in \$/MWh, of which the units are dropped hereafter for simplicity. As concept demonstrations,  these small setups, although not realistic in practice, are sufficiently complex to reveal non-trivial characteristics of multi-interval dispatch and pricing.


\subsection{Simulation settings}
\begin{figure}[h]
\center
\begin{psfrags}
\scalefig{0.5}\epsfbox{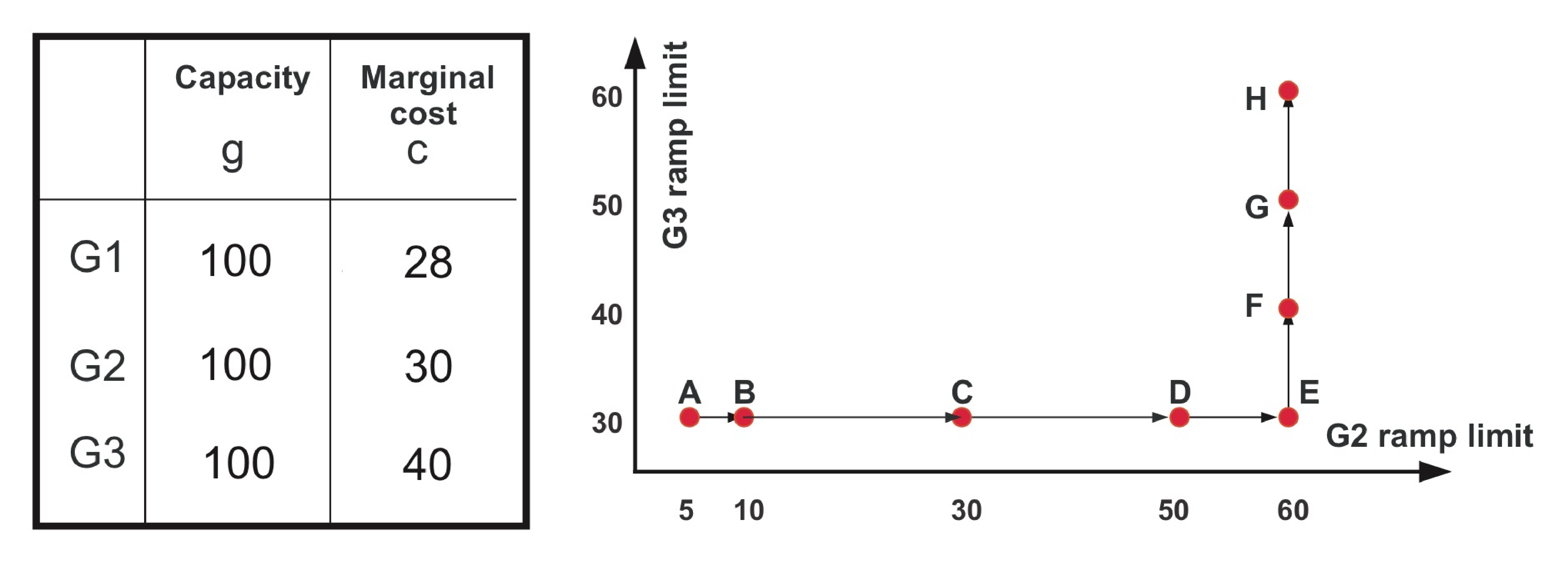}
\scalefig{0.24}\epsfbox{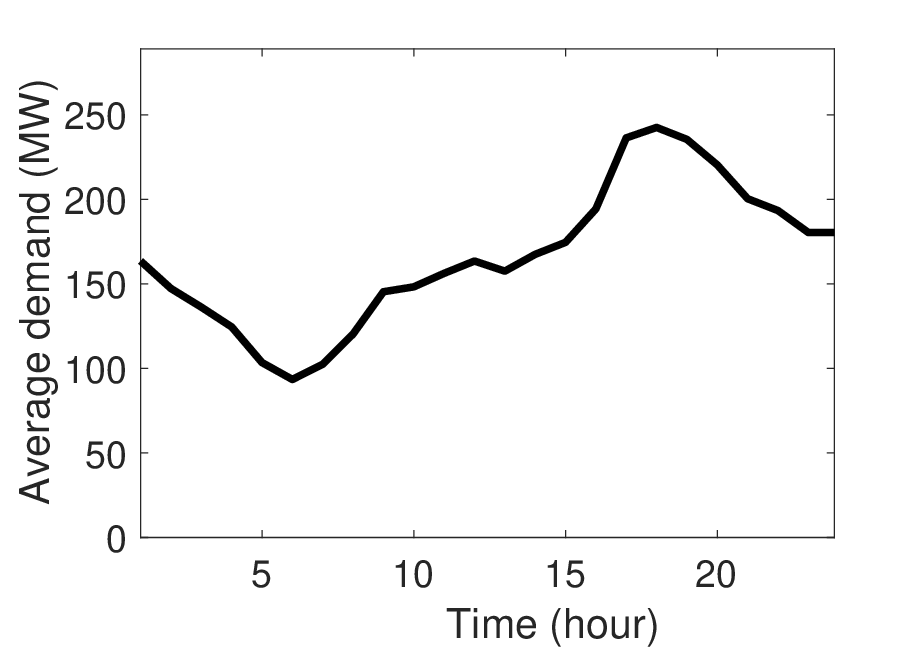}\scalefig{0.24}\epsfbox{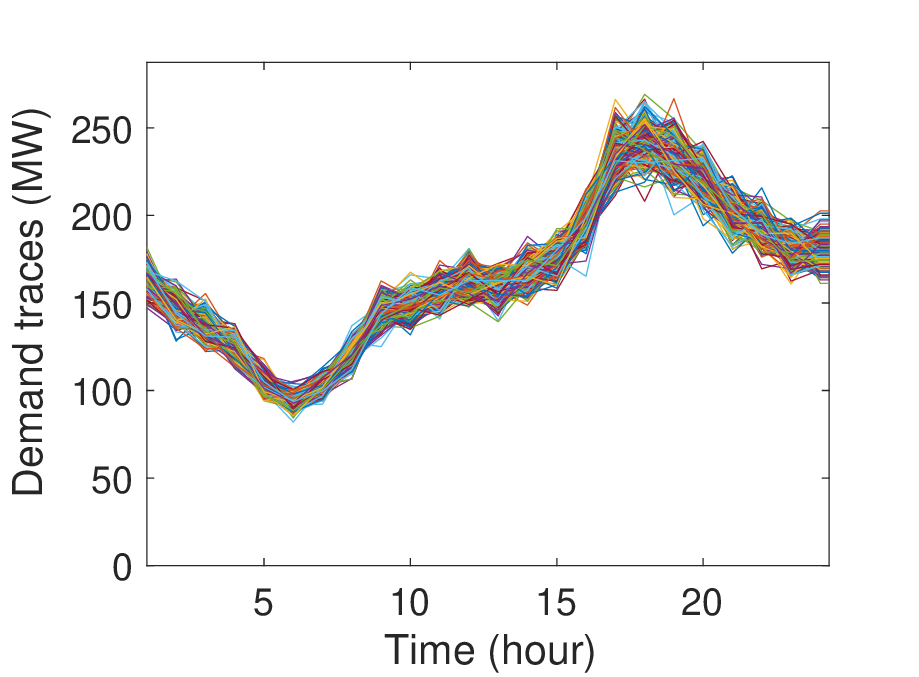}
\end{psfrags}
\vspace{-1em}\caption{\small Top left: generator parameters.  The ramp limit for G1 is fixed at 25 (MW/h).  Top right: a path of ramping events. Bottom left: average demand.  Bottom right: demand traces.}
\label{fig:demand}
\end{figure}

The top part of Fig~\ref{fig:demand}  shows the parameters of the generators  and a ramping path used in the simulations.  Specifically, we evaluated the performance of benchmark schemes by varying ramping limits of G2 and G3 along the path from scenario A to H while fixing the ramping limit of generator G1 to 25 MW/h. Scenario A had the most stringent ramping constraints and H the most relaxed.

The bottom part of Fig~\ref{fig:demand} shows the 300 realizations and average demand over 24 hour period generated from a CAISO load profile and  a standard deviation of $4\%$ of the mean value.   We used a standard forecasting error model\footnote{The forecast $\hat{d}_{(t+k)|t}$ at $t$  of demand $d_{t+k}$ is $\hat{d}_{(t+k)|t}=d_{t+k}+\sum_{i=1}^k \epsilon_k$ where $\epsilon_k$ is i.i.d. Gaussian with zero mean and variance $\sigma^2$.} where the demand forecast $\hat{d}_{(t+k)|t}$ of $d_{t+k}$  at time $t$ had error variance $k\sigma^2$  increasing linearly with $k$.

All simulations were conducted with rolling-window optimization over the 24-hour scheduling period, represented by 24 time intervals. And the window size is four intervals in each rolling window optimization.

\subsection{Dispatch-following and ramping-revelation incentives}

\subsubsection{LOC and dispatch-following incentives}  We first considered dispatch-following incentives measured by the LOC payment; the greater the LOC payment, the higher the incentive to deviate the dispatch signal (in the absence of LOC payment).  The computation of LOC for the pricing models followed that defined in Part I of the paper and given in detail in the appendix of this paper.

{\scriptsize
\begin{figure}[h]
\center
\begin{psfrags}
\scalefig{0.25}\epsfbox{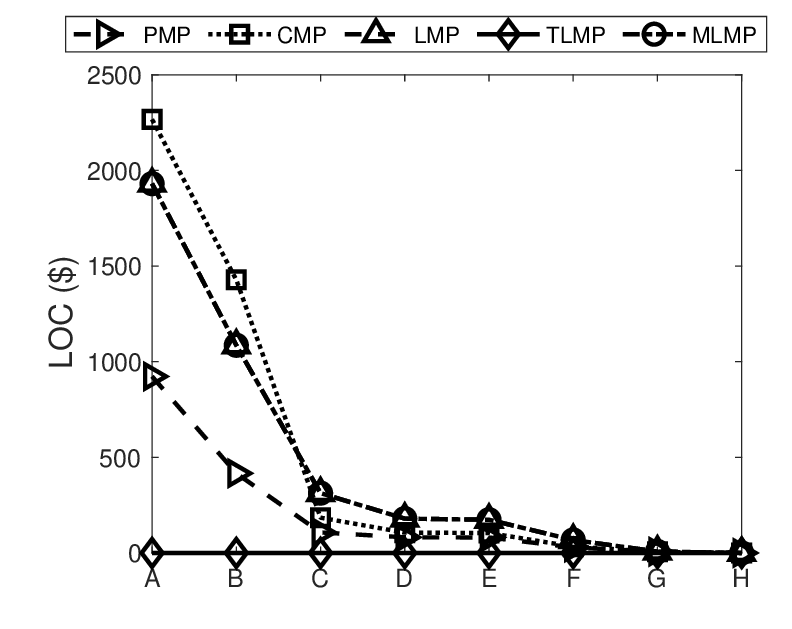}\scalefig{0.25}\epsfbox{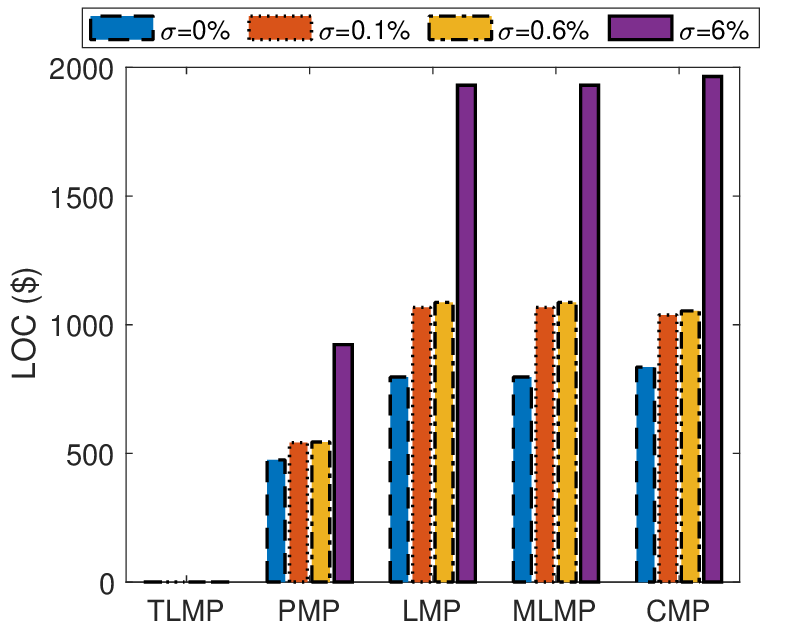}
\end{psfrags}
\vspace{-2em}
\caption{\scriptsize Left panel: LOC vs. ramping scenarios from A to H at $\sigma=6\%$. Right panel: LOC for ramping scenario A under forecast error standard deviation $\sigma=0\%, 0.1\%, 0.6\%, 6\%$. }
 \label{fig:LOC}
\end{figure}
}

Fig.~\ref{fig:LOC} shows the total LOC payment from the ISO to generators at different ramping rates along the ramping trajectory in Fig.~\ref{fig:demand}. Notice the general trend that all schemes converged to zero as scenarios of binding ramping constraints diminished at scenario H.

As predicted by the equilibrium property, the LOC for TLMP was strictly zero, and all other pricing schemes had positive LOC payments.  PMP designed to minimize the LOC appeared to have the least LOC among the rest of the uniform pricing schemes. The same conclusion held for the larger scale simulations considered in the appendix.  Shown also in Fig.~\ref{fig:LOC} is that LOC increased with the forecasting error variance, as expected.

\subsubsection{Truthful revelation of ramping limits}  This simulation aimed at illustrating incentives of the truthful revelation of ramping limits under various pricing schemes.  We varied the {\em revealed} ramping limit of one generator and kept the others fixed at the true ramp limits.

\begin{figure}[h]
\center
\begin{psfrags}
\scalefig{0.16}\epsfbox{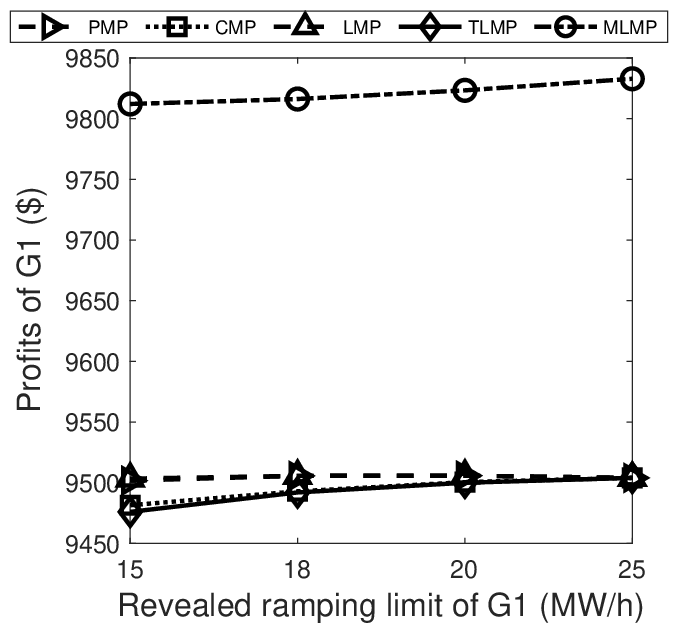}\scalefig{0.16}\epsfbox{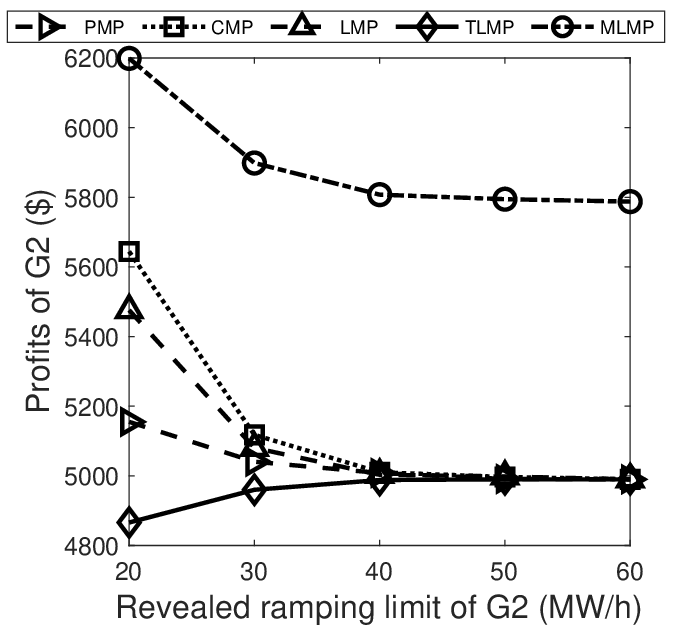}\scalefig{0.16}\epsfbox{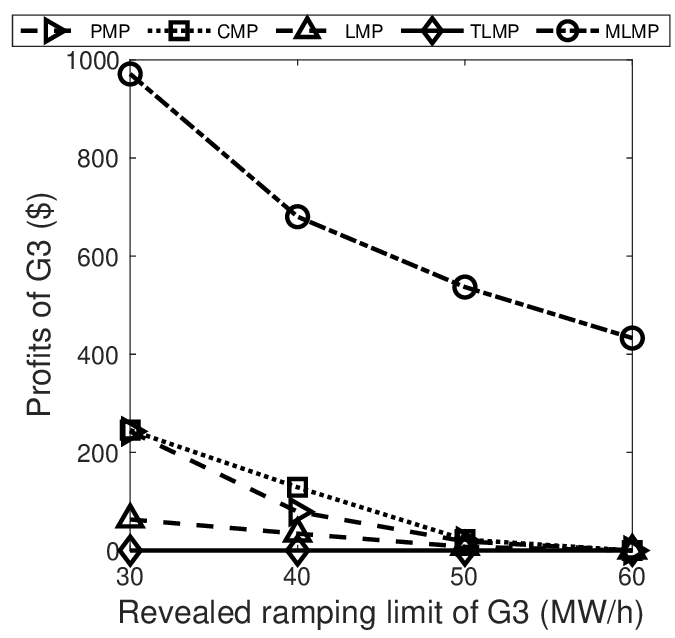}
\end{psfrags}
\vspace{-0em}\caption{\scriptsize Generator profit vs. revealed ramp limit at $\sigma=6\%$ for ramp scenario H.    Left: ramp limits of G2 and G3 are fixed at 60 MW/h.  Middle: ramp limit of G1 is fixed at 25 MW/h and G3 at 60 MW/h.  Right: ramp limits of G1 is fixed 25 MW/h and G2 at 60 MW/h. }
 \label{fig:Revelation}
\end{figure}

Fig.~\ref{fig:Revelation} shows the generator profit as a function of its {\em revealed ramping limits} for the ramping scenario H with true ramp limits as 25 MW/h for G1, 60 MW/h for G2 and G3. Under TLMP,  profits of all generators grew as the revealed ramping limits grew  to their true values.  The implication was that the generators had incentives to reveal their ramp limits truthfully and to improve their ramping capabilities.   For the rest of uniform pricing schemes, the profits of generators G2 and G3  increased as the revealed ramp limits deviate  from their true values, implying that generators had incentives to under-report their ramp limits.

\begin{figure}[h]
\center
\begin{psfrags}
\scalefig{0.16}\epsfbox{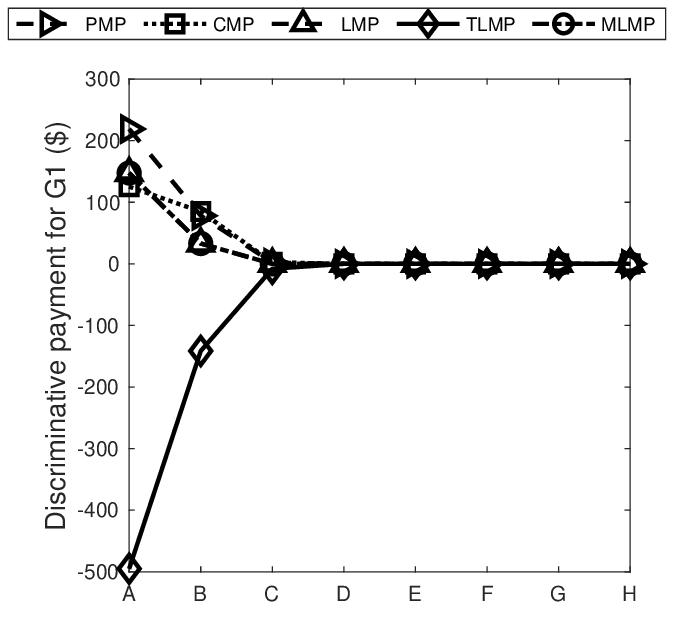}\scalefig{0.16}\epsfbox{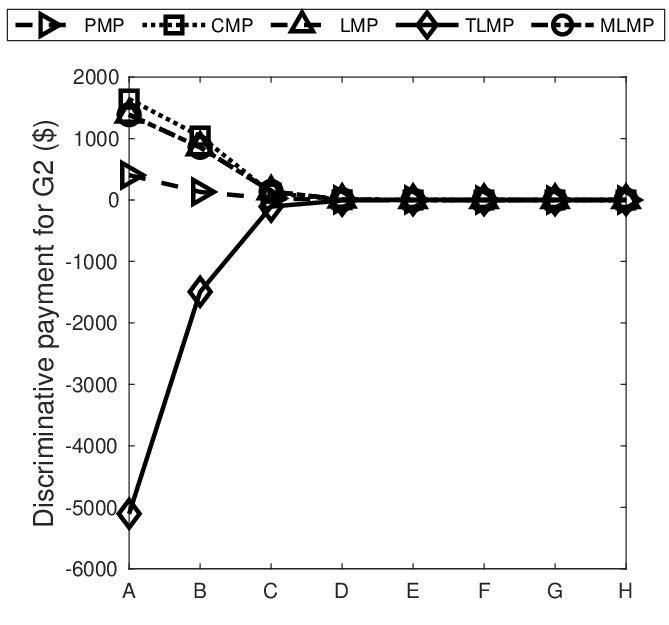}\scalefig{0.16}\epsfbox{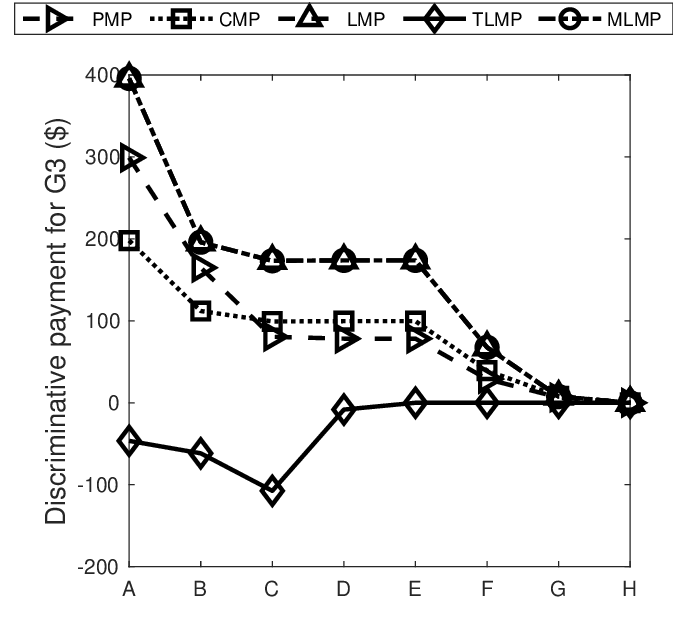}
\end{psfrags}
\vspace{-0em}\caption{\scriptsize Discriminative payment vs. ramping scenarios from A to H at $\sigma=6\%$. Left: Discriminative payment of G1. Middle: Discriminative payment of G2. Right: Discriminative payment of G3. }
 \label{fig:Discrimination}
\end{figure}

\subsection{In-market and out-of-market discriminative payment} Fig.~\ref{fig:Discrimination} shows the comparison of discriminative payments to different generators under different pricing schemes. The discriminative payments to each generator under uniform prices (LMP, PMP, MLMP and CMP) equaled to the corresponding LOC, which were also called out-of-market discriminative payments. And the discriminative payment under TLMP was in-market discrimination, which was calculated by the payment to generator under TLMP minus that under LMP. Note that the discriminative payment of TLMP is mostly negative, indicating that generators with binding ramping constraints tend to be paid lower than LMP. Comparing to other uniform pricing schemes, the generator with more consecutive binding ramping constraints, i.e. G3 in Fig.~\ref{fig:Discrimination}, had smaller discriminative payment under TLMP. While generators with more nonconsecutive binding ramping constraints, i.e. G1 and G2, had more discriminative payments under TLMP.

Shown also in Fig.~\ref{fig:Discrimination} is that there's no certain order for the absolute values of discriminative payments under different pricing schemes. 

\subsection{Revenue adequacy of ISO}
\begin{figure}[h]
\center
\begin{psfrags}
\scalefig{0.25}\epsfbox{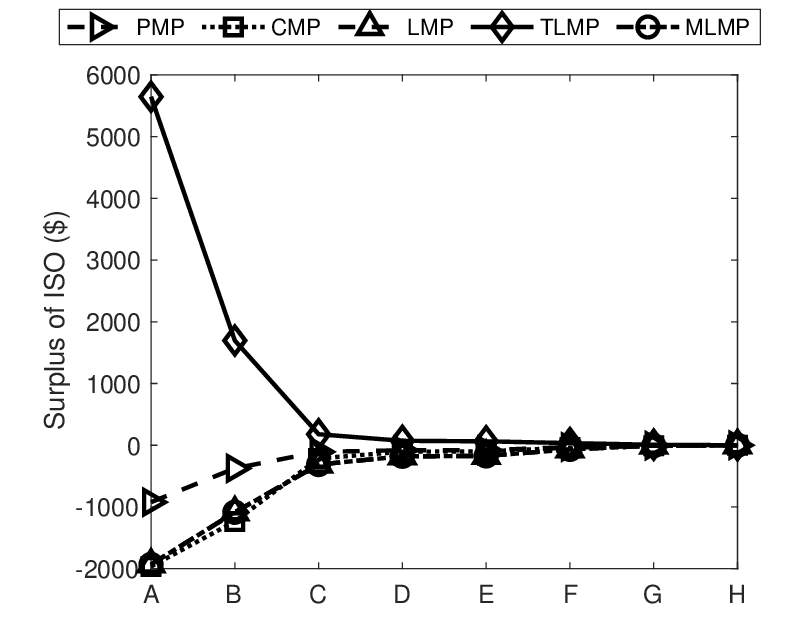}\scalefig{0.25}\epsfbox{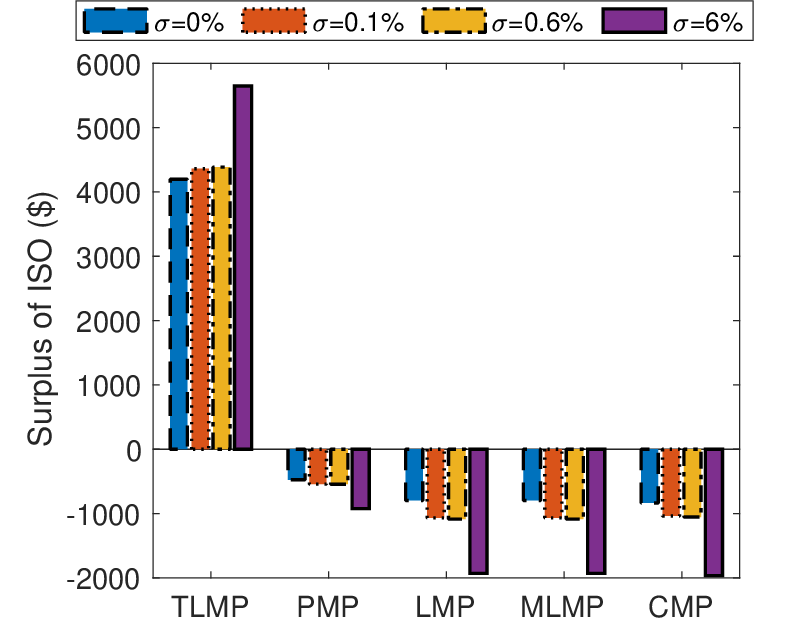}
\end{psfrags}
\vspace{-2em}\caption{\scriptsize ISO surplus vs. ramp limits.  Left panel: ISO surplus evaluated at $\sigma=6\%$.  Right panel: Ramping scenario A.}
 \label{fig:ISOrevenue}
\end{figure}

Fig.~\ref{fig:ISOrevenue} shows the ISO's merchandising  surplus that included the LOC payments.  The results validated the fact that uniform pricing schemes, in general, have positive LOC, resulting in a deficit for the ISO.  As a regulated utility, any deficit (and surplus) was redistributed to the consumers in a revenue reconciliation process \cite{Schweppe&Caramanis&Tabors&Bohn:88book}.

For TLMP, the ramping charge on generators led to a positive merchandising surplus, as shown in Proposition~\ref{prop3:RevenueTLMP}.  The simulations involving a larger network in the appendix also  showed that the rolling-window TLMP had a merchandising surplus from both ramping and congestion.
  Coupled with the fact that TLMP always had zero LOC, TLMP showed a positive merchandising surplus.

The ISO surpluses for all pricing schemes converged to the congestion rent (which was zero in the single-node case)
as ramping events diminished with increasing ramping limits.   For TLMP, the ISO surplus decreased from the positive because ISO collected less penalty charges from generators.  The ISO surpluses for all other pricing schemes increased from the negative because of the decreasing LOC payments.

\subsection{Consumer payments and generator profits}  We assumed that ISO was financially neutral;   when the ISO had a positive surplus (after excluding the congestion surplus),  the consumers received a price reduction as a rebate.  When the ISO had a deficit, the consumers paid additionally to cover the deficit.

\begin{figure}[h]
\center
\begin{psfrags}
\scalefig{0.25}\epsfbox{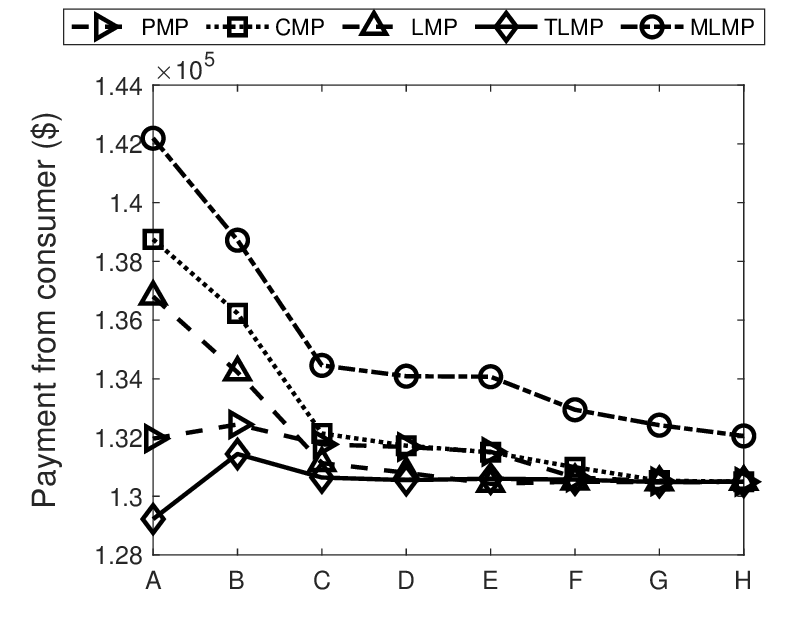}\scalefig{0.25}\epsfbox{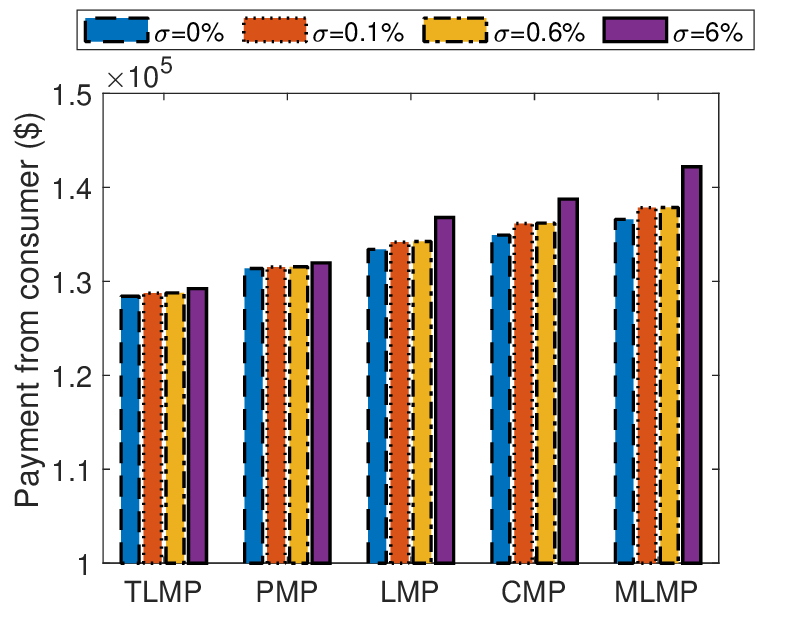}
\end{psfrags}
\vspace{-2em}\caption{\scriptsize Consumer payment   vs. ramp.  Left panel: consumer payment evaluated at $\sigma=6\%$.  Right panel: Ramping scenario A.}
 \label{fig:ConsumerPay}
\end{figure}

 Fig.~\ref{fig:ConsumerPay} shows the consumer payments under the assumption that the demand is credited (or charged) for any ISO surplus (or deficit).   TLMP was the least expensive for the consumer and PMP the least expensive among uniform pricing schemes.   The decreasing trend of consumer payments with less ramping constraints under  uniform pricing schemes  was due to the decreasing costs of  LOC payments to the generators.  The initial increasing trend of consumer payment under TLMP was due to the less surplus of ISO passed to the consumers for collecting penalties from generators.    Again, the consumer payments increased with the forecasting error.

The total generator profit figures have identical trends as those of consumer payments because the operator has zero surplus.  TLMP had the least generator profits, and PMP had the least generator profits among uniformly priced schemes.
Note that the forecasting errors resulted in higher generator profits for LMP, CMP, and MLMP because of high LOC payments to generators.

\begin{figure}[hbt]
\center
\begin{psfrags}
\scalefig{0.4}\epsfbox{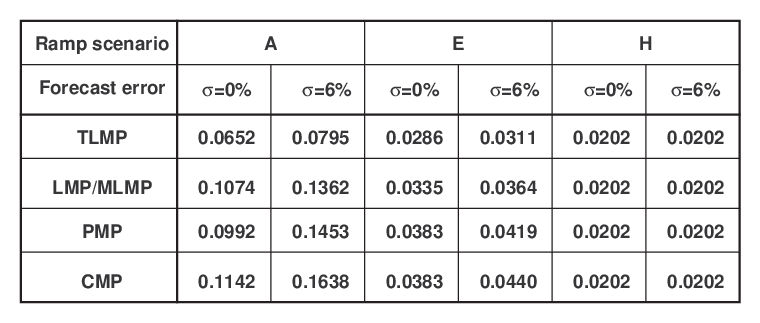}
\scalefig{0.25}\epsfbox{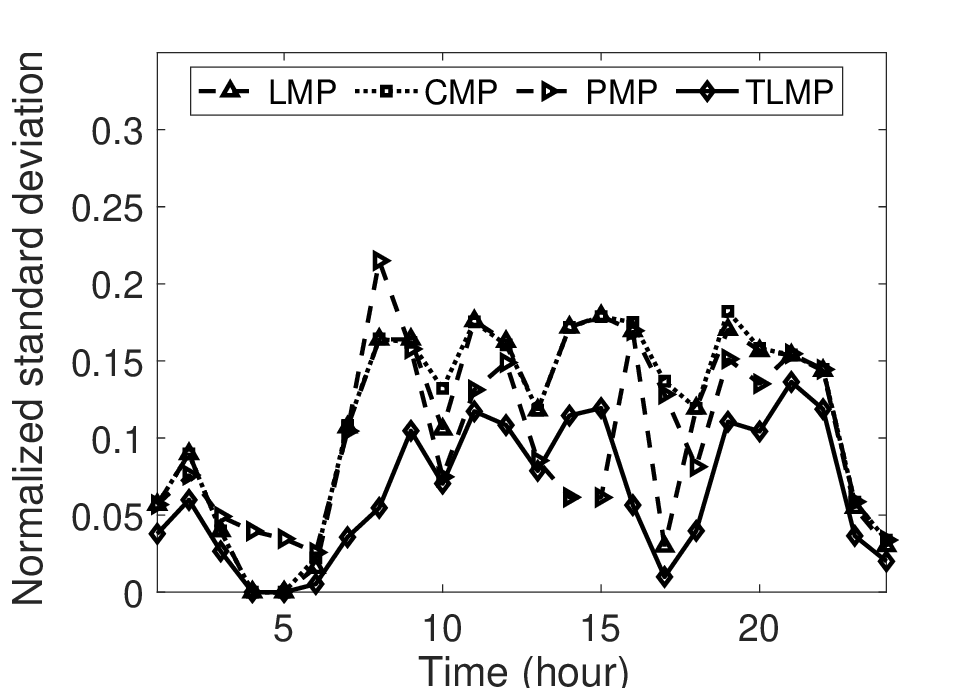}\scalefig{0.25}\epsfbox{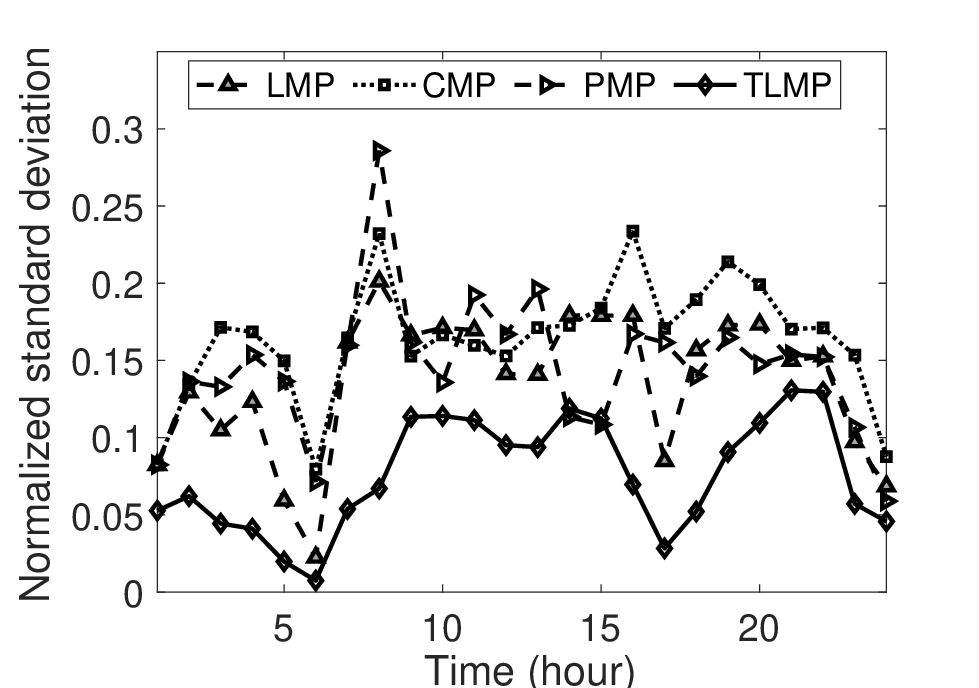}
\end{psfrags}
\vspace{-2em}\caption{\scriptsize Average ratio of normalized standard deviation  of hourly prices. Top: normalized standard deviation under different ramping scenarios and standard deviation of forecasting errors.  Bottom: normalized standard deviation at different hours with $\sigma=0\%$ (left) and $\sigma=6\%$  (right).}
 \label{fig:PriceSTD}
\end{figure}

\subsection{Price volatility}
The volatility of a random price in an hour can be measured by the standard deviation of the price normalized by the average of the price in the hour.  A highly volatile price makes LMP forecasting difficult.

Fig.~\ref{fig:PriceSTD} includes a table of price volatility averaged over all hours.  Among the compared pricing mechanisms, TLMP\footnote{The normalized standard deviation of TLMP is averaged over for all the demand and generators.} showed consistently lower volatility.  We also noticed that  price volatility increased with stricter ramping limits and
increasing demand forecasting errors.  The same trend was also observed in simulations involving larger networks in the appendix of this paper.

\section{Conclusion}\label{sec:conclusion}
This two-part paper considers the pricing of multi-interval dispatch under demand forecast uncertainty.  We establish that, to provide dispatch-following incentives, discrimination in the form of uniform pricing with out-of-the-market uplifts or nonuniform pricing becomes necessary.   In particular, we show that, as a generalization of LMP,  the nonuniform TLMP eliminates the need for the out-of-the-market uplifts under arbitrary forecasting uncertainty.  We also consider incentives of the truthful revelation of ramping limits.  We show that, by penalizing the ramping limits, TLMP provides incentives for generators to improve its ramping capability and reveal the actual ramping limits.  Unfortunately, such incentives are lacking in the existing pricing schemes.

Under the rolling-window dispatch, different pricing schemes differ in the distribution of the overall social welfare among generators and consumers.   Among the pricing mechanisms considered in this paper, TLMP leads to the least consumer payment but also the lowest generator profit.  Likewise, among uniform pricing schemes, PMP leads to the least consumer payment and the lowest generator profits.

%
\section*{Acknowledgement}
The authors are grateful for the many discussions with Dr. Tongxin Zheng whose insights helped to shape this two-part paper.  We are also benefited from helpful comments and critiques from Shumel Oren, Kory Hedman, Mojdeh Abdi-Khorsand, Timothy Mount, and Bowen Hua.
{
\bibliographystyle{IEEEtran}
\bibliography{BIB}

\begin{thebibliography}{10}
\providecommand{\url}[1]{#1}
\csname url@samestyle\endcsname
\providecommand{\newblock}{\relax}
\providecommand{\bibinfo}[2]{#2}
\providecommand{\BIBentrySTDinterwordspacing}{\spaceskip=0pt\relax}
\providecommand{\BIBentryALTinterwordstretchfactor}{4}
\providecommand{\BIBentryALTinterwordspacing}{\spaceskip=\fontdimen2\font plus
\BIBentryALTinterwordstretchfactor\fontdimen3\font minus
  \fontdimen4\font\relax}
\providecommand{\BIBforeignlanguage}[2]{{%
\expandafter\ifx\csname l@#1\endcsname\relax
\typeout{** WARNING: IEEEtran.bst: No hyphenation pattern has been}%
\typeout{** loaded for the language `#1'. Using the pattern for}%
\typeout{** the default language instead.}%
\else
\language=\csname l@#1\endcsname
\fi
#2}}
\providecommand{\BIBdecl}{\relax}
\BIBdecl

\bibitem{Guo&Tong:18Allerton}
Y.~{Guo} and L.~{Tong}, ``Pricing multi-period dispatch under uncertainty,'' in
  \emph{2018 56th Annual Allerton Conference on Communication, Control, and
  Computing (Allerton)}, Oct 2018, pp. 341--345.

\bibitem{Tong:19PESGM}
L.~{Tong}, ``Pricing multi-period dispatch under uncertainty,'' in \emph{Proc.
  2019 IEEE PES General Meeting}, August 2019.

\bibitem{Hogan:16}
W.~Hogan, ``Electricity market design: Optimmization and market equilibrium,''
  [ONLINE], available (2019/9/9) at
  \url{https://sites.hks.harvard.edu/fs/whogan/Hogan_UCLA_011316.pdf}, January
  2016.

\bibitem{Hua&etal:19TPS}
B.~{Hua}, D.~A. {Schiro}, T.~{Zheng}, R.~{Baldick}, and E.~{Litvinov},
  ``Pricing in multi-interval real-time markets,'' \emph{IEEE Transactions on
  Power Systems}, vol.~34, no.~4, pp. 2696--2705, July 2019.

\bibitem{Zhao&Zheng&Litvinov:19arxiv}
J.~Zhao, T.~Zheng, and E.~Litvinov, ``A multi-period market design for markets
  with intertemporal constraints,'' [ONLINE], available (2019/9/9) at
  \url{https://arxiv.org/abs/1812.07034}, June 2019.

\bibitem{Guo&Chen&Tong:19arxiv}
Y.~Guo, C.~Chen, and L.~Tong, ``Pricing multi-interval dispatch under
  uncertainty: Part {I}---dispatch-following incentives,'' [ONLINE] at
  \url{https://arxiv.org/abs/1911.05784}, October 2020.

\bibitem{Thatte&Choi&Xie:14PSCC}
A.~A. {Thatte}, D.~{Choi}, and L.~{Xie}, ``Analysis of locational marginal
  prices in look-ahead economic dispatch,'' in \emph{2014 Power Systems
  Computation Conference}, Aug 2014, pp. 1--7.

\bibitem{Wu&Variaya&Spiller&Oren:96JRE}
F.~Wu, P.~Varaiya, P.~Spiller, and S.~Oren, ``Folk theorems on transmission
  access: Proofs and counter examples,'' \emph{Journal of Regulatory
  Economics}, vol.~10, no.~1, pp. 5--23, July 1996.

\bibitem{Schiro:17}
D.~A. Schiro, ``Procurement and pricing of ramping capability,'' [ONLINE],
  available (2019/9/9) at
  \url{https://www.iso-ne.com/static-assets/documents/2017/09/20170920-procurement-pricing-of-ramping-capability.pdf},
  September 2017.

\bibitem{Schweppe&Caramanis&Tabors&Bohn:88book}
F.~C. Schweppe, M.~Caramanis, R.~D. Tabors, and R.~E. Bohn, \emph{Spot pricing
  of electricity}.\hskip 1em plus 0.5em minus 0.4em\relax New York: Kluwer
  Academic Publishers, 1988.

\bibitem{krishnamurthy:15TPS}
D.~Krishnamurthy, W.~Li, and L.~Tesfatsion, ``An 8-zone test system based on
  iso new england data: Development and application,'' \emph{IEEE Transactions
  on Power Systems}, vol.~31, no.~1, pp. 234--246, 2015.

\bibitem{Krishnamurthy:15Bit}
D.~Krishnamurthy, ``8-zone iso-ne test system: Code and data repository,''
  [ONLINE], available (2015/4/29) at
  \url{https://bitbucket.org/kdheepak/eightbustestbedrepo/src/master/}, April
  2015.

\end{thebibliography}
}

\section*{Appendix}
\subsection{Proof of Proposition~1}
Let $\qbf^{\RED}[t]$ be the vector of injections in interval $t$  from the R-ED policy.  From (\ref{eq:RLMP}), the ISO surplus  is given by
\bea
\mbox{\rm MS}^{\RLMP} &:=& \sum_{t=1}^T (\dbf[t]-\qbf^{\RED}[t])^{\intercal}(\lambda_t^{\RLMP} {\bf 1} - \Sbf^{\intercal} \pmb{\phi}^{\RLMP}[t])\nn\\
&=& \sum_{t=1}^T  (\qbf^{\RED}[t]-\dbf[t])^{\intercal}  \Sbf^{\intercal} \pmb{\phi}^{\RLMP}[t]\nn\\
&=&  \sum_{t=1}^T   \cbf^{\intercal} \pmb{\phi}^{\RLMP}[t],  \nn
\eea
where the last equality uses the complementary slackness condition of (\ref{eq:RED}).  \hfill\QED

\subsection{Proof of Proposition~2}
The proof of Proposition~2 follows directly from the Proof of Proposition~2 in Part I. See \cite{Guo&Chen&Tong:19arxiv}.

 \subsection{Proof of Proposition~3}
 The proof of Proposition 3 follows that of Proposition~1.

 \subsection{Proof of Proposition~4}
  Because Proposition~4 focuses on the payment to generator $i$ at bus $m$, we drop the subscripts $i$ and superscript $m$ in the relevant notations.
  \bea
  \Delta R &:=& R^{\LMP}-R^{\TLMP} \nn\\
  &=& \sum_{t=1}^T (\pi_t^{\LMP}-\pi^{\TLMP})g_t^*\nn\\
  &= & \sum_{t=1}^T \Delta \mu^*_{t-1}g_t^* - \sum_{t=1}^T \Delta\mu_t^*g_t^*\nn\\
  &=& \sum_{t=1}^T \Delta \mu^*_{t-1}(g_t^*-g_{t-1}^*)  +  \Delta\mu^*_{0}g_{0}^* -  \Delta\mu^*_T g^*_{T} \nn\\
  &=&  \sum_{t=1}^T \Delta \mu^*_{t-1}(g_t^*-g_{t-1}^*). \nn
  \eea
  By the complementary slackness condition,
  \bea
  \Delta \mu^*_{t}(g_{t+1}^*-g_{t}^*) &=& \left\{\begin{array}{ll}
  \bar{\mu}^*_t\bar{r}_i & g_{t+1}^*-g_{t}^* = \bar{r}\\
   \underline{\mu}^*_t\underline{r}_i &  g_{t+1}^*-g_{t}^* =-\underline{r}\\
   0 & \mbox{otherwise}\\
   \end{array}
   \right.\nn\\
 &=& \bar{\mu}_t^*\bar{r}+ \underline{\mu}_t^*\underline{r} \ge 0. \nn
   \eea
   \hfill\QED

   \subsection{Proof of Proposition~5}  The proof of Proposition~5 follows exactly the proof of Theorem~5 given in Appendix F of \cite{Guo&Chen&Tong:19arxiv}.

\subsection{Pricing optimization of PMP and CMP}
\subsubsection{Optimization in PMP}  The rolling-window PMP policy $\Gc^{\RPMP}_t$ solves the following optimization:
\beq \label{eq:PMPa}
\begin{array}{lrl}
 & \Gc^{\RPMP}_t:& \mbox{at time $t$,} \\[0.2em]
&\underset{\{\Gbf\}}{\rm minimize}   &  F(\Gbf) -\sum_{t'=1}^{t-1} \qbf^{\intercal}[t']\pibf^{\RPMP}[t']
\\
&  \mbox{subject to:} & \mbox{Network constraints: }\\
& & \qbf[t']=(\sum_i g_{it'}^1,\cdots,\sum_i g_{it'}^M),\\
& \lambda_{t'}: & {\bf 1}^{\intercal} \qbf[t'] ={\bf 1}^{\intercal} \hat{\dbf}[t']\\
&\pmb{\phi}[t']: & \Sbf (\qbf[t']-\hat{\dbf}[t']) \le \cbf\\
& & \mbox{for all $t\le t' < t+W$.}\\[0.2em]
 &  & \mbox{Generation constraints:}\\
& (\underline{\mu}^m_{it'},\bar{\mu}^m_{it'}):  &  -\underline{r}^m_i\le g^m_{i(t'+1)}-g^m_{it'} \le \bar{r}^m_{i},\\
& (\underline{\rho}^m_{it'},\bar{\rho}^m_{it'}):   & 0 \le g^m_{it'} \le \bar{g}^m_{i}\\
 & &  \mbox{for all $m$, $0 < t' < t+W$.}
\end{array}
\eeq
The rolling-window PMP sets the price for generation in interval $t$ by
\beq \label{eq:PMPa1}
\pibf^{\RPMP}[t] = \lambda^{\RPMP}_t {\bf 1}- \Sbf^{\intercal}\pmb{\phi}^{\RPMP}[t],
\eeq
where $\lambda^{\RPMP}_t$ and $\pmb{\phi}^{\RPMP}[t]$ are the multipliers
associated with power balance and line-flow constraints in (\ref{eq:PMPa}).

\subsubsection{Optimization in CMP}
Let $\Gbf=\big[\gbf[t],\cdots, \gbf[t+W-1]\big]$ be the generation variables within the $W$-interval lookahead window, and $F_t(\Gbf)$ the total cost of generation.
 The rolling-window CMP policy $\Gc_t^{\RCMP}$ solves the following optimization:
\beq \label{eq:RCMPa}
\begin{array}{lrl}
 & \Gc^{\RCMP}_t:& \mbox{at time $t$,}\\[0.2em]
 &\underset{\{\Gbf\}}{\rm minimize}   &  F_t(\Gbf)  + \sum_{m,i} (\bar{\mu}_{mit}^{\RED}-\underline{\mu}_{mit}^{\RED}) g^m_{it}\\
 &  \mbox{subject to:} & \mbox{Network constraints:}\\
 & & \qbf[t']=(\sum_i g_{it'}^1,\cdots,\sum_i g_{it'}^M),\\
& \lambda_{t'}: & {\bf 1}^{\intercal} \qbf[t'] ={\bf 1}^{\intercal} \hat{\dbf}[t']\\
&\pmb{\phi}[t']:  & \Sbf (\qbf[t']-\hat{\dbf}[t']) \le \cbf\\
 & & \hfill \mbox{for all $t\le t'\ < t+W$.}\\[0.2em]
 &  & \mbox{Generation constraints:}\\
& &  -\underline{r}^m_i\le g^m_{i(t'+1)}-g^m_{it'} \le \bar{r}^m_{i},\\
&  & 0 \le g^m_{it'} \le \bar{g}^m_{i}, \\
 & & \hfill \mbox{for all  $m, t\le t' < t+W$.}\\[0.2em]
   &  & \mbox{boundary ramping constraints:}\\
  & &  -\underline{r}^m_i\le g^m_{it}-g^{\RED}_{mi(t-1)} \le \bar{r}^m_{i}.\\
\end{array}
\eeq

\subsection{Computation of LOC}
\subsubsection{Single settlement prices}
 For the single settlement pricing schemes, the computation of LOC is with respect to a specific generator under dispatch vector $\gbf=(g_1,\cdots,g_T)$ and price vector $\pibf=(\pi_1,\cdots,\pi_T)$.  The LOC associated with $\pibf$ is given by
 \beq \label{eq:LOC}
 \mbox{LOC}(\pibf) = Q(\pibf) - (\pibf^{\intercal}\gbf - F(\gbf))
 \eeq
where  $F(\gbf)=\sum_t f_t(g_t)$ is the total (true) cost of generating $\gbf$, and $Q(\pibf) $ is the post-uplift  generator-surplus defined by
\beq \label{eq:Q}
\begin{array}{lll}
Q(\pibf)= & \underset{\pbf=(p_1,\cdots, p_T)}{\rm maximize} & \sum_{t=1}^T(\pi_{t}p_t - f_{t}(p_t)) \\[0.5em]
& {\rm subject~to}&  \underline{g}\leq p_{t} \leq \bar{g}\\
&  & -\underline{r}\leq p_{(t+1)}-p_t \leq \bar{r}. \\
\end{array} \hfill
\eeq
Using (\ref{eq:LOC}-\ref{eq:Q}),
the computation of LOC under the rolling-window version of LMP, TLMP, PMP,and CMP are made by substituting $\gbf$ by the rolling-window dispatch $\gbf^{\RED}$ and $\pibf$ by the rolling-window of the corresponding  prices. \hfill\QED

\subsubsection{Multi-settlement  prices}
For the multi-settlement pricing such as MLMP, note that the generator can only affect the  revenue by setting the dispatch in the binding interval.  Therefore,  in calculating $\mbox{LOC}^{\MLMP}$, the only decision variables are the realized
generations.

Let $\tilde{g}_t$ be the pre-binding dispatch of interval $t$, \ie, $\tilde{g}_t$ is the advisory dispatch level $\hat{g}_{it}^{m,W-1}$ for generator $i$ at bus $m$ in the $(W-1)$th settlement as shown in Fig.~\ref{fig:MS}.  The generator profit maximization problem is given by
\beq \label{eq:Q1}
\begin{array}{lll}
Q(\pibf^{\MLMP})= & \underset{\pbf=(p_1,\cdots, p_T)}{\rm maximize} & \sum_{t=1}^T(\pi^{\RLMP}_{t}(p_t-\tilde{g}_{t}) - f_{t}(p_t)) \\[0.5em]
& {\rm subject~to}&  \underline{g}\leq p_{t} \leq \bar{g}\\
&  & -\underline{r}\leq p_{(t+1)}-p_t \leq \bar{r}. \\
\end{array} \hfill\nn
\eeq
Since $\tilde{g}_{t}$ is given, the above optimization has the same solution as $\pbf^*$ as that of (\ref{eq:Q}) for R-LMP using $\pibf^{\RLMP}$ (although the values of the two optimization are different).

The LOC for MLMP is therefore
\bea
\mbox{LOC}^{\MLMP}
&=& \sum_{t=1}^T \bigg(\pi^{\RLMP}_{t}(p^*_t-g_{t(t-1)}) - f_{t}(p^*_t)\bigg) \nn\\
& &  -    \bigg(\sum_{t=1}^T \pi^{\RLMP}_{t}(g^{\RED}_t-g_{t(t-1)})  - F(\gbf^{\RED})\bigg)\nn\\
&=& \mbox{LOC}^{\RLMP}    \ge 0. \nn
\eea

\begin{figure}[h]
\center
\begin{psfrags}
\scalefig{0.48}\epsfbox{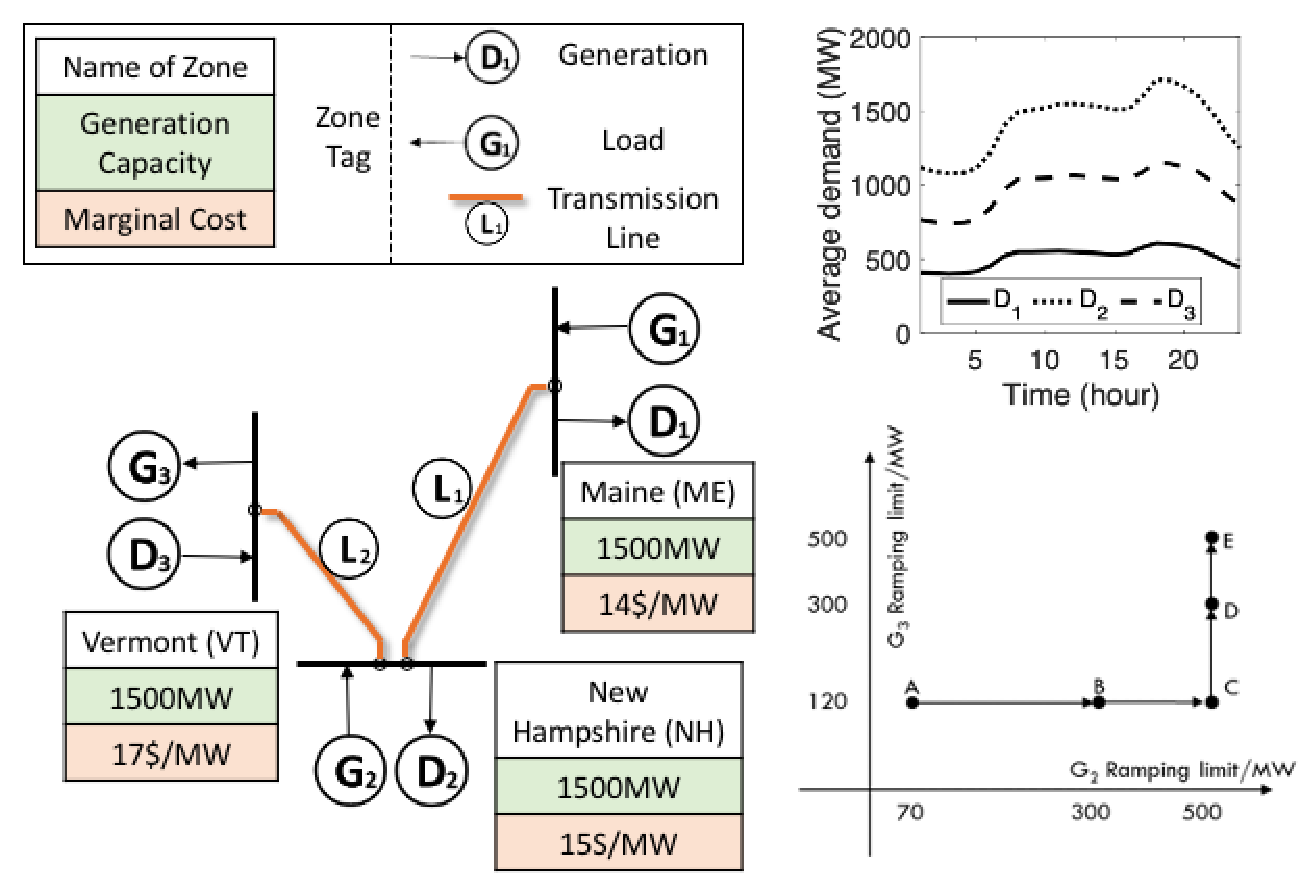}
\end{psfrags}
\caption{\scriptsize  Left panel: the 3-node-2-transmission line case. Right panel: average load demand and ramping scenarios (The ramping limit for G1 is fixed at 300MW/h).}
\label{fig:networksetting}
\end{figure}

\subsection{Simulation results for a system with network constraints} \label{sec:network}
We considered a three-generator two-transmission line network abstracted from the 8-zone ISO New England (ISO-NE) case shown in Fig.~\ref{fig:networksetting}. Base on an ISONE load data profile, 300 load scenarios were generated as in the single bus case.   The standard deviation of the one-step forecasting error was set at $\sigma=0.6\%$.  Other settings in these simulations are similar to the one-node case unless otherwise specified.

Pricing performance was evaluated with varying ramping limits. Line capacity of Line 1 (L1) was 1000MW, and Line 2 (L2) was always congestion-free. The ramp limits of generations G2 and G3 varied along the path from scenario A to E with A having the most stringent ramping constraints and E the most relaxed.

Conclusions drawn for the single-node cases held mostly in simulations involving a network with power flow constraints.  We provide here additional comments, especially related to congestions.

\begin{figure}[h]
\centering
\begin{psfrags}
\scalefig{0.25}\epsfbox{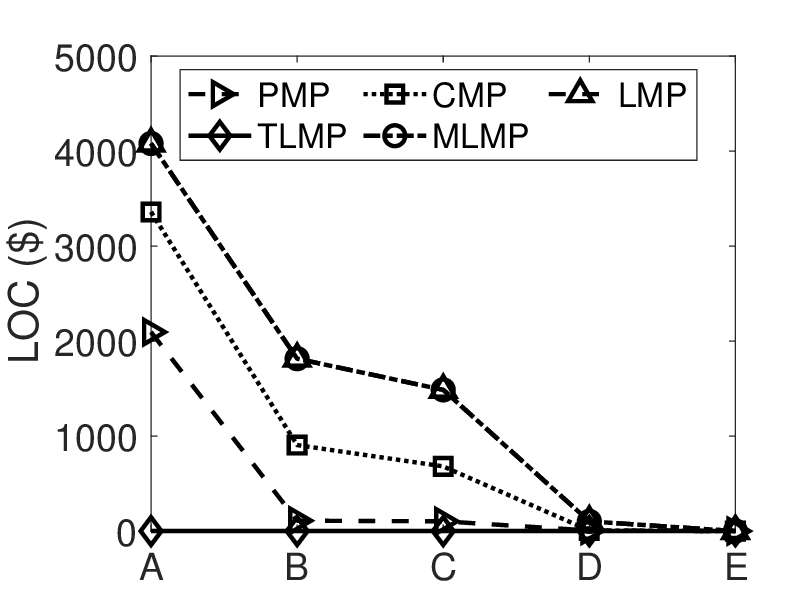}\scalefig{0.25}\epsfbox{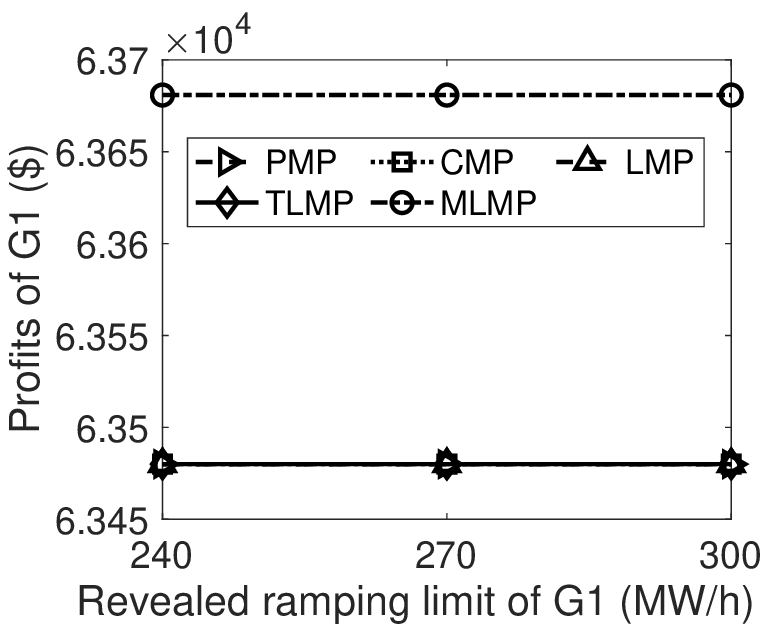}\\
\scalefig{0.25}\epsfbox{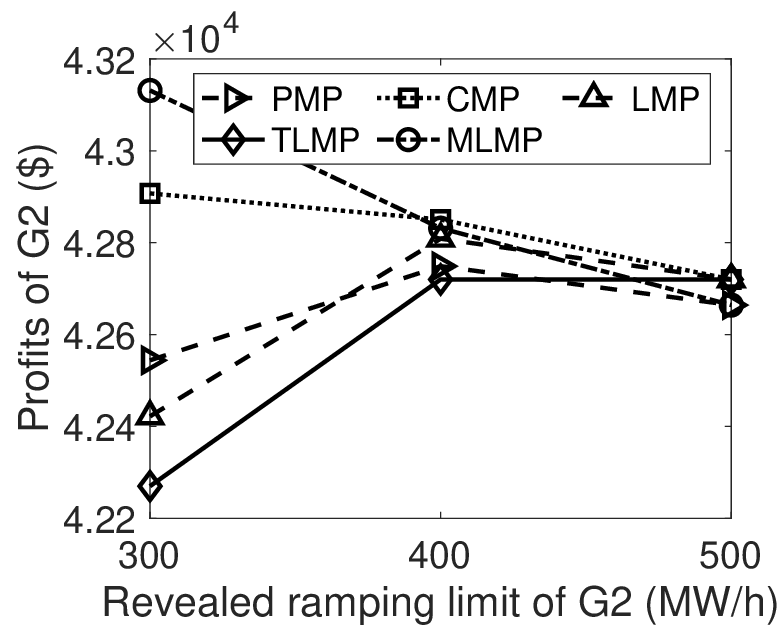}\scalefig{0.25}\epsfbox{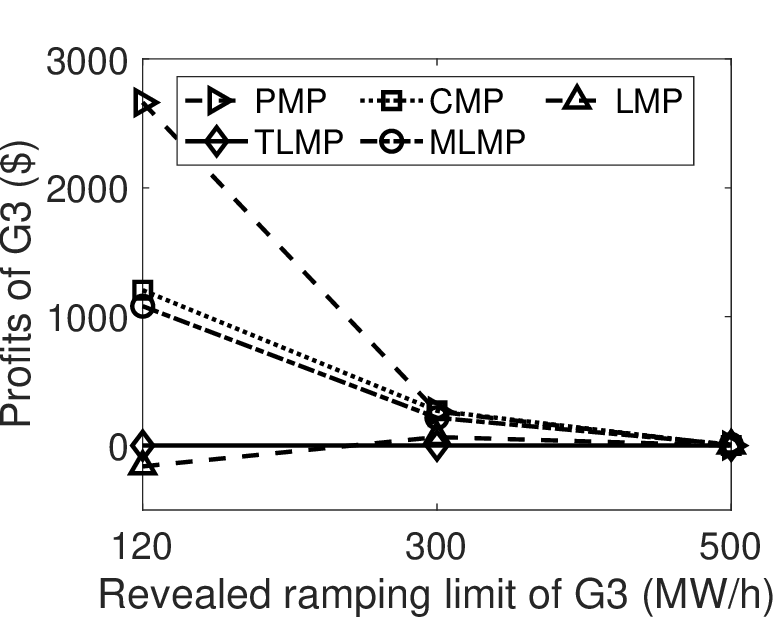}\\
\end{psfrags}
\vspace{-0.75em}\caption{\scriptsize  Top left: LOC vs. ramping scenarios from A to E. The other figures are individual generator profit vs. its revealed ramp limit.  The true ramp limits are 300 MW/h for G1, 500 MW/h for G2 and G3.}
\label{fig:Network_LOCvsRamp}
\end{figure}

\subsubsection{Dispatch-following and ramping-revelation incentives}
Fig.~\ref{fig:Network_LOCvsRamp} shows measures of incentives.  The top left panel of Fig.~\ref{fig:Network_LOCvsRamp} shows the total LOC payment at different ramping rates along the ramping trajectory in Fig.~\ref{fig:networksetting}.     As shown in Proposition~\ref{prop5:EQTLMP},  the strong equilibrium property of TLMP dictated that its LOC should be strictly zero.   All other pricing schemes had LOC with PMP having the least amount.  The LOC converged when ramping constraints vanished at scenario E.

In evaluating incentives for generators to reveal its ramping limits truthfully, we fixed all but one generator at their true ramp limits and varied the revealed ramping limits of a  single generator.  The three figures in Fig.~\ref{fig:Network_LOCvsRamp} show that only TLMP received the highest profit when the revealed ramp limit matched with the actual value.  There are incentives for G2 and G3 to under-report ramping limits for all other pricing schemes.

\subsubsection{Revenue adequacy of ISO }
The revenue adequacy  ISO  when the network is in congestion needs to take into account congestion rent. The total ISO surplus is thus given by
\bea
\mbox{ISO surplus} &=& \mbox{Payment from demand} - \mbox{Congestion rent} \nn\\
& & -\mbox{Payment to generators (inc LOC)}.\nn
\eea

\begin{figure}[h]
\center
\begin{psfrags}
\scalefig{0.25}\epsfbox{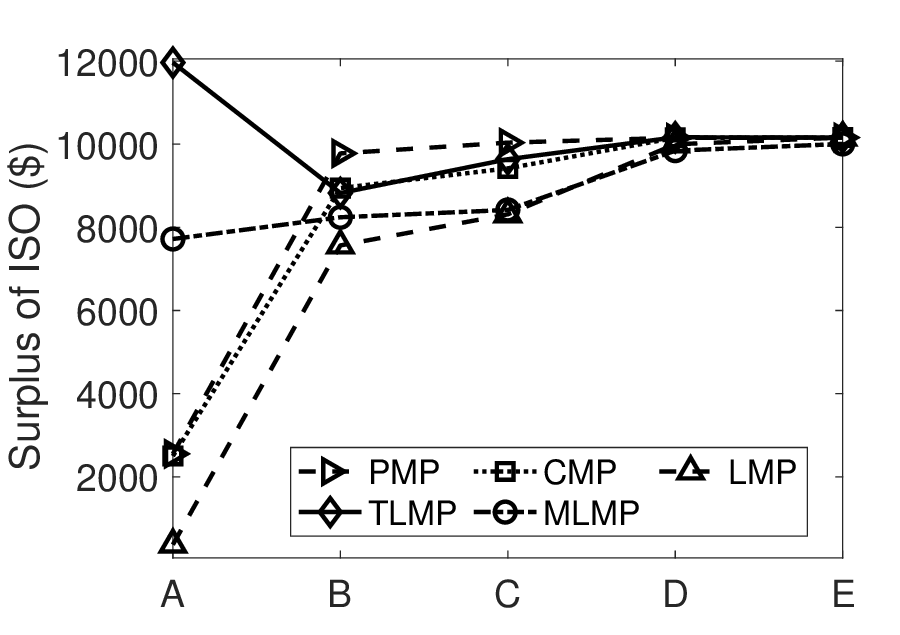}\scalefig{0.25}\epsfbox{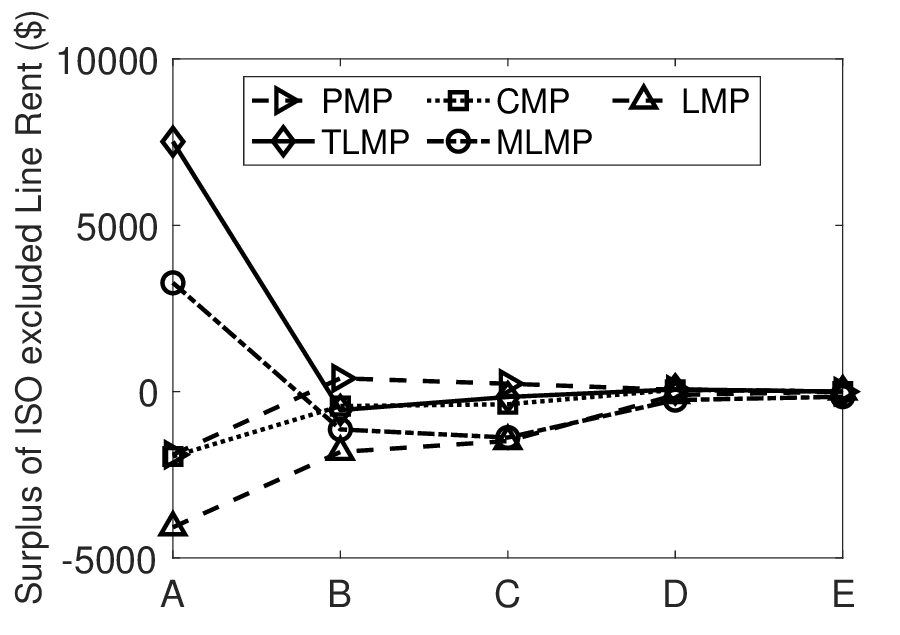}
\end{psfrags}
\vspace{-2em}\caption{\small  Left: ISO surplus without excluding congestion rent. Right: ISO surplus after excluding congestion rent.}
 \label{fig:Network_ISOsurplusvsRamp}
\end{figure}

Fig.~\ref{fig:Network_ISOsurplusvsRamp}  shows the surplus of ISO without excluding (left) congestion rent and after excluding (right) congestion rent.   Before the  congestion rent was
excluded, all pricing mechanisms showed a positive surplus for ISO, indicating that demand payments include substantial congestion-related payments.   Note also that all pricing schemes converged to the congestion rent as ramping limits diminished under scenario E.

After congestion rent was removed, uniform pricing schemes had a noticeable negative ISO surplus.  Under TLMP, ISO remained to be revenue adequate for all cases except under scenario C where there was a small deficit.   Note that this is not in contradiction to Proposition~\ref{prop3:RevenueTLMP} where the one-shot dispatch and perfect forecasting are assumed.

\subsubsection{Consumer payments and generator profits}
\begin{figure}[h]
\center
\begin{psfrags}
\scalefig{0.25}\epsfbox{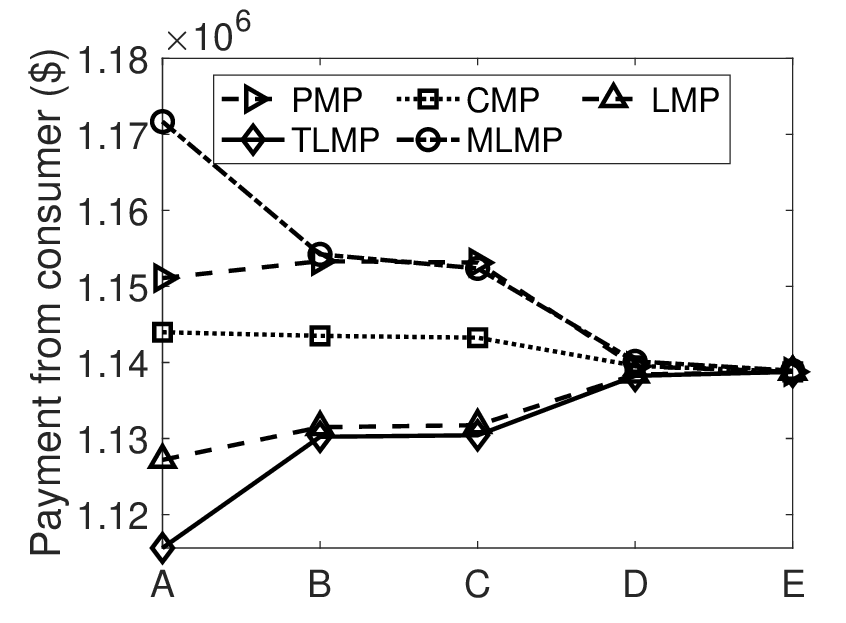}\scalefig{0.25}\epsfbox{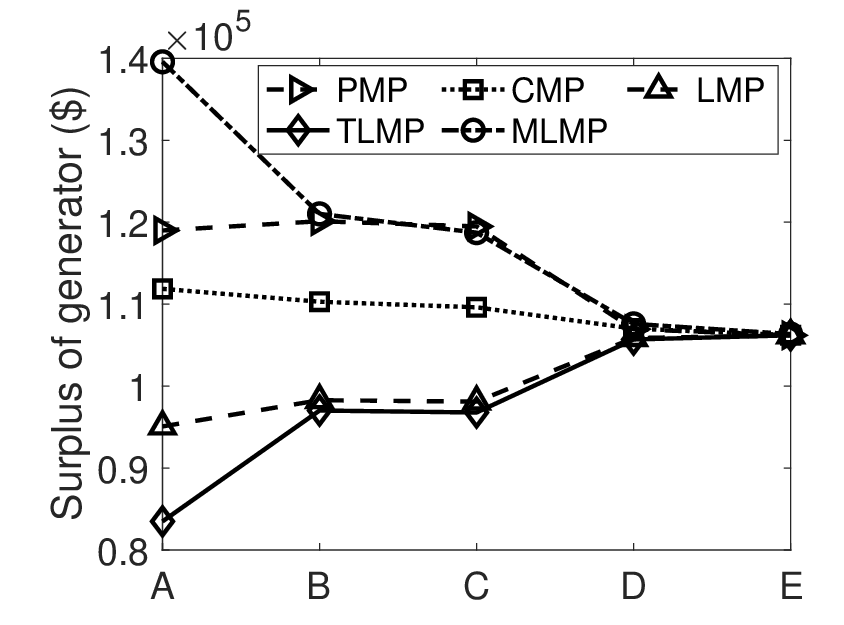}
\end{psfrags}
\vspace{-1em}\caption{\small  Left: consumer payment vs ramping limits. Right: surplus of generators vs ramping limits.}
 \label{fig:Network_ConsumerPayvsRamp}
\end{figure}

Fig~\ref{fig:Network_ConsumerPayvsRamp} shows the consumer payments (left) and generator profits (right)  under the assumption that the operator charges its shortfall (and returns its profit) to consumers.  Note that consumer payments and generator profits were strongly dependent; the lower the consumer payment was, the lower generator payments were.

Consumer payments under TLMP were shown to be the lowest.  Unlike the single-node case, LMP had the least consumer payment among uniform pricing schemes. Note also that, except for LMP, all uniform pricing schemes resulted in decreasing consumer payments when ramping constraints were relaxed from A to E.  The consumer payment for TLMP increased along the same path of ramping scenarios.   These trends were consistent with the single-node case.  The trend for LMP, however, did not follow that in the single-node case due to a complicated interaction of congestion and ramping constraints.

%

\begin{figure}[h]
\center
\begin{psfrags}
\scalefig{0.4}\epsfbox{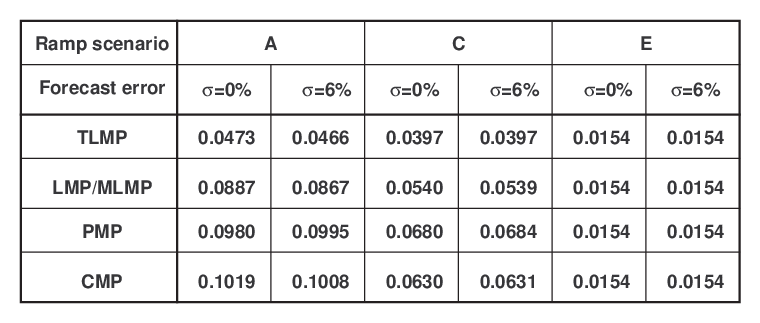}
\scalefig{0.25}\epsfbox{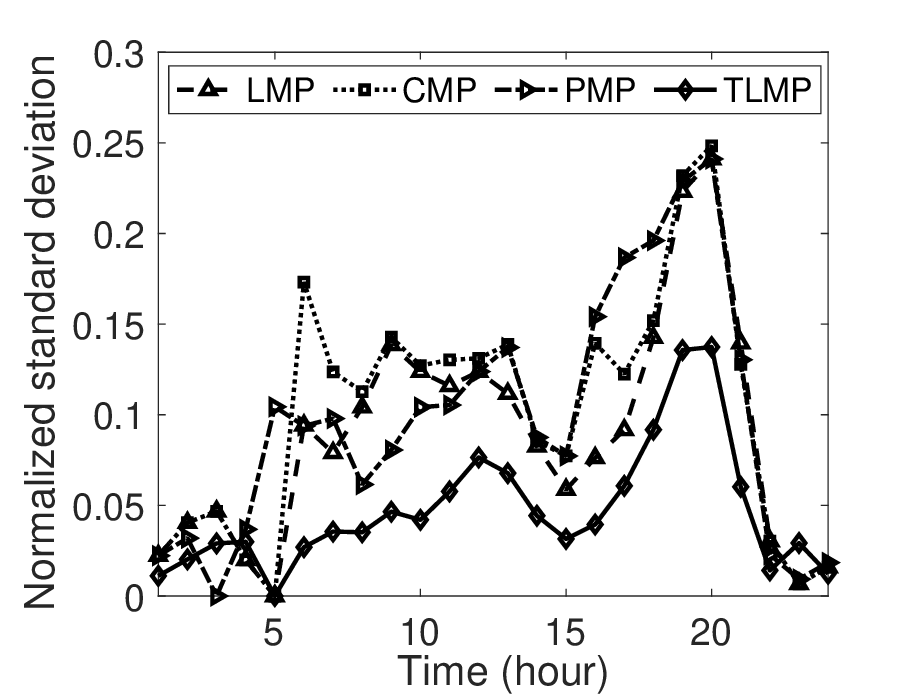}\scalefig{0.25}\epsfbox{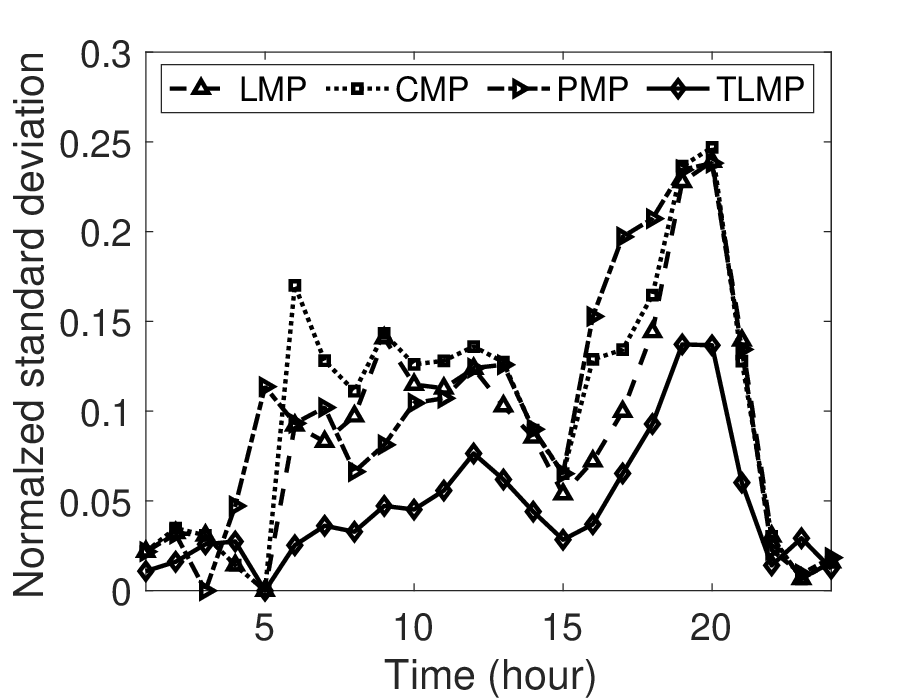}
\end{psfrags}
\vspace{-0.5em}\caption{\small Average ratio of normalized standard deviation  of hourly prices. Top: normalized standard deviation under different ramping scenarios and standard deviation of forecasting errors.  Bottom: normalized standard deviation at different hours with $\sigma=0\%$ (left) and $\sigma=6\%$  (right).}
 \label{fig:Network_PriceSTD}
\end{figure}

\subsubsection{Price volatility}
The price volatility was evaluated under both ramping and congestion constraints.
Fig.~\ref{fig:Network_PriceSTD}  shows that the average price standard deviation table under strict (A), relaxed (C), and unconstrained (E) ramping limit scenarios.  The price standard deviation in the table was averaged over the 24-hour period whereas the two figures below are the hourly average price standard deviation.

The same conclusions as in the single node case held. For the most part, TLMP appeared to be the least volatile among pricing schemes evaluated in this paper.

\begin{figure}[h]
\center
\begin{psfrags}
\scalefig{0.5}\epsfbox{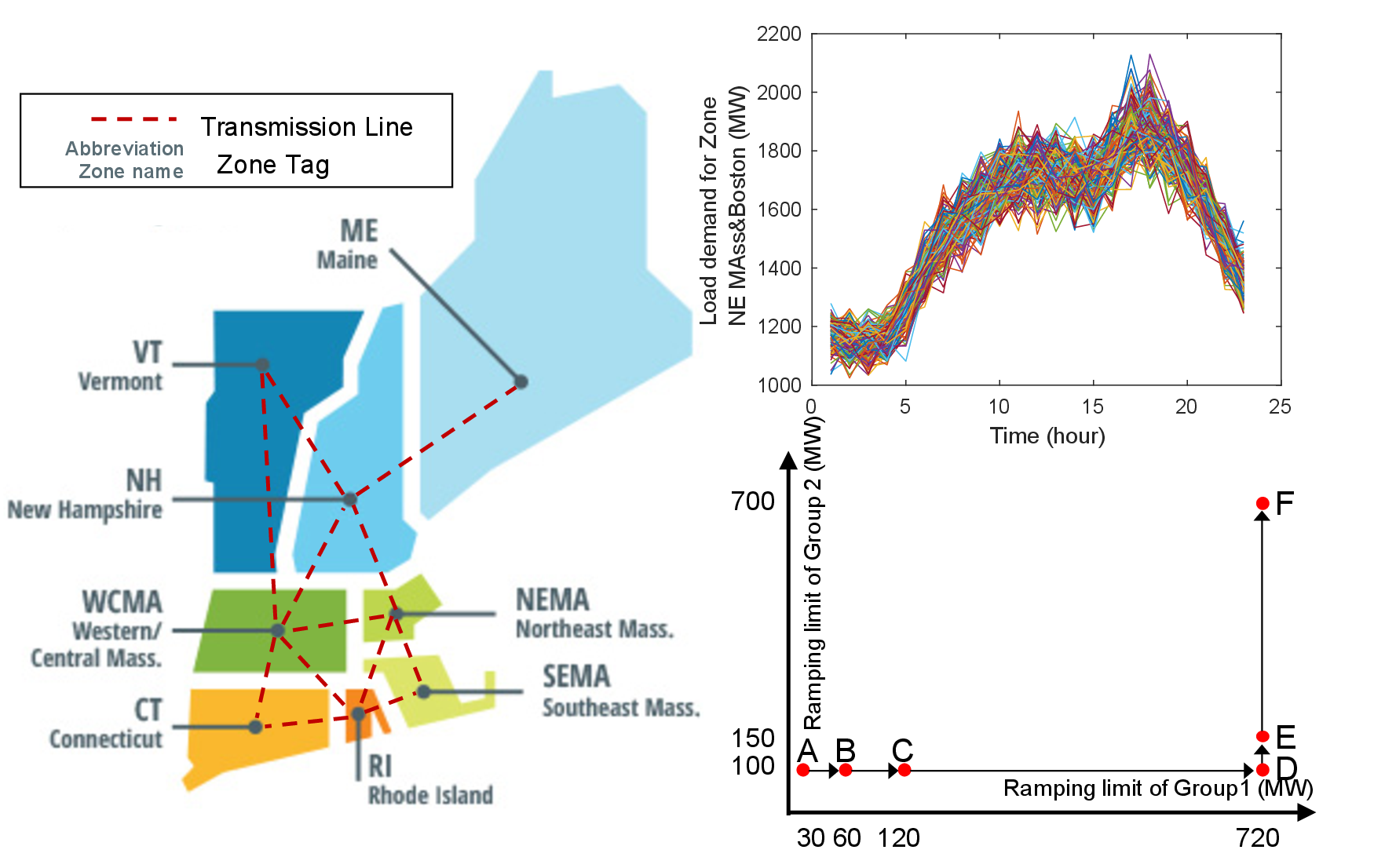}
\end{psfrags}
\vspace{-1em}\caption{\small  Left: ISO-NE system. Right: Load scenarios at NE Mass \& Boston and ramping scenarios}
\label{fig:ISONEsetting}
\end{figure}

\subsection{Simulation results for the 76 generator ISO-NE system} \label{sec:ISO-NE network}
We considered the 8-zone ISO New England (ISO-NE) case with 76-generator and 12-transmission line shown in Fig.~\ref{fig:ISONEsetting} with parameters from   \cite{krishnamurthy:15TPS,Krishnamurthy:15Bit}. All line capacities were set at 1000MW. Monte Carlo simulations of 200 load scenarios were conducted with normalized standard deviation of the one-step forecasting error set at $\sigma=0.6\%$. Other settings in these simulations were similar to the one-node case unless otherwise specified.

Pricing performance was evaluated with varying ramping limits. In the ISO-NE  data set, there were two groups of generators with different ramping limits. The ramping limit of Group 1 varied from 30MW/h to 120MW/h and that of Group 2 varied from 100MW/h to 700MW/h. From Fig.~\ref{fig:ISONEsetting}, the ramp limits of generators in two groups varied along the path from scenario A to F with A having the most stringent ramping constraints and F the most relaxed.

Conclusions drawn from previous single-node cases and three-generator two-transmission line cases mostly held for the ISO-NE simulations.  Here we  highlight some of the differences in the ISO-NE test cases from the previous smaller-scale simulations.

\begin{figure}[h]
\centering
\begin{psfrags}
\scalefig{0.5}\epsfbox{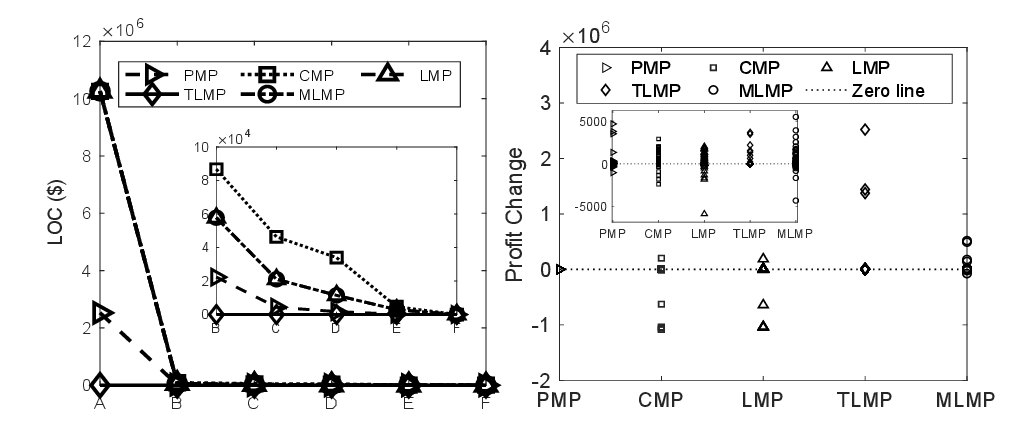}
\end{psfrags}
\vspace{-0.75em}\caption{\scriptsize  Left: LOC vs. ramping scenarios from A to F.  Right: under scenario A, the profit change of one generator  when its ramp limits  was changed by 10\%.}
\label{fig:ISONE_LOCvsRamp}
\end{figure}

\subsubsection{Dispatch-following and ramping-revelation incentives}
As is shown on the left panel of Fig.~\ref{fig:ISONE_LOCvsRamp}, only TLMP  supported dispatch-following incentives with strictly zero LOC, as expected from  Proposition~\ref{prop5:EQTLMP}. This simulation also showed that the average LOC decreased dramatically from ramping setting A to B where  ramp limits of  generators in Group 1 increased  from 30MW to 60MW. See also the performance in the magnified LOC plot for the ramping path from B to F.

For the truthful revelation of ramping limits, the right panel of Fig.~\ref{fig:ISONE_LOCvsRamp} shows the scatter plot of the profit change of a generator when its ramp limits was increased by 10\% while others were truthful.  When the profit change was negative, it meant that the the generator had incentive not to reveal the true ramp limit.  Among the five benchmark pricing schemes, TLMP had the smallest number of   non-incentive compatible cases.

\begin{figure}[h]
\center
\begin{psfrags}
\scalefig{0.5}\epsfbox{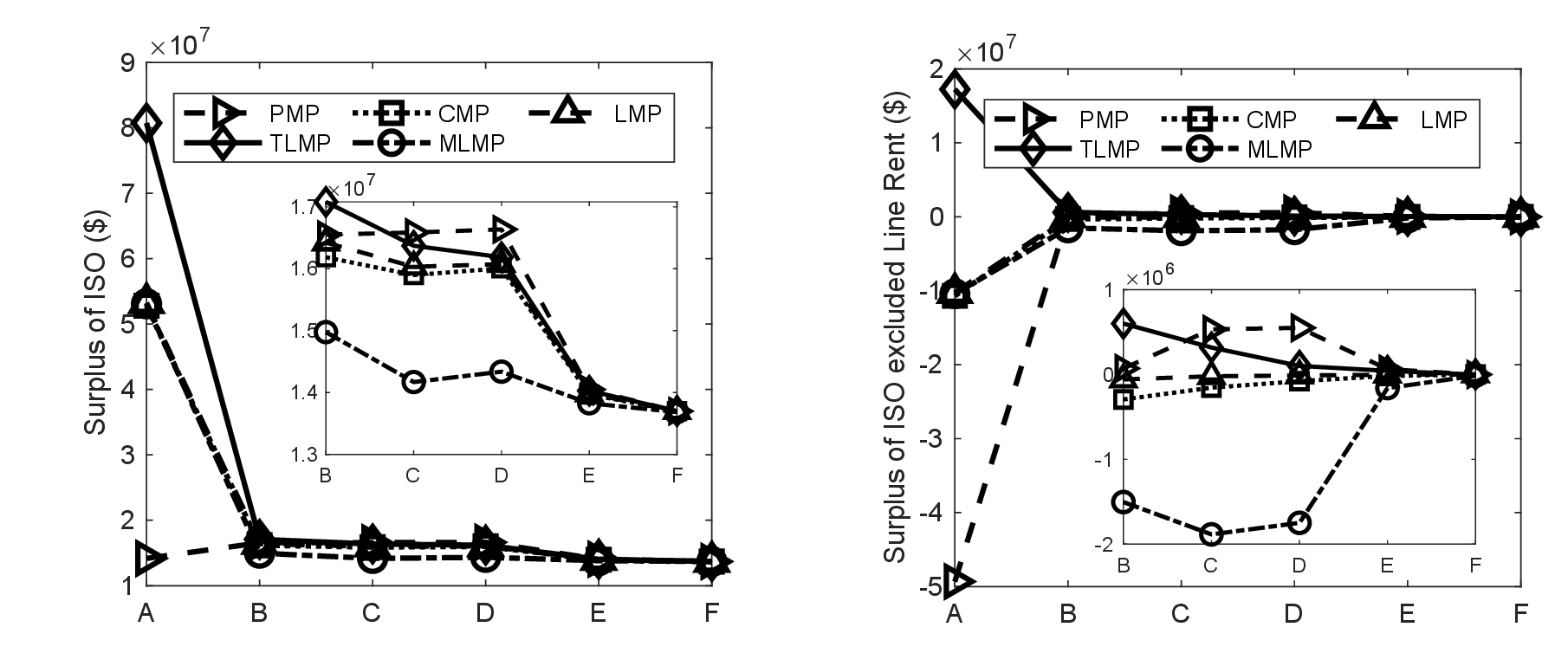}\\
\end{psfrags}
\vspace{-1em}\caption{\small  Left: ISO surplus without excluding congestion rent. Right: ISO surplus after excluding congestion rent.}
 \label{fig:ISONE_ISOsurplusvsRamp}
\end{figure}

\begin{figure}[h]
\center
\begin{psfrags}
\scalefig{0.5}\epsfbox{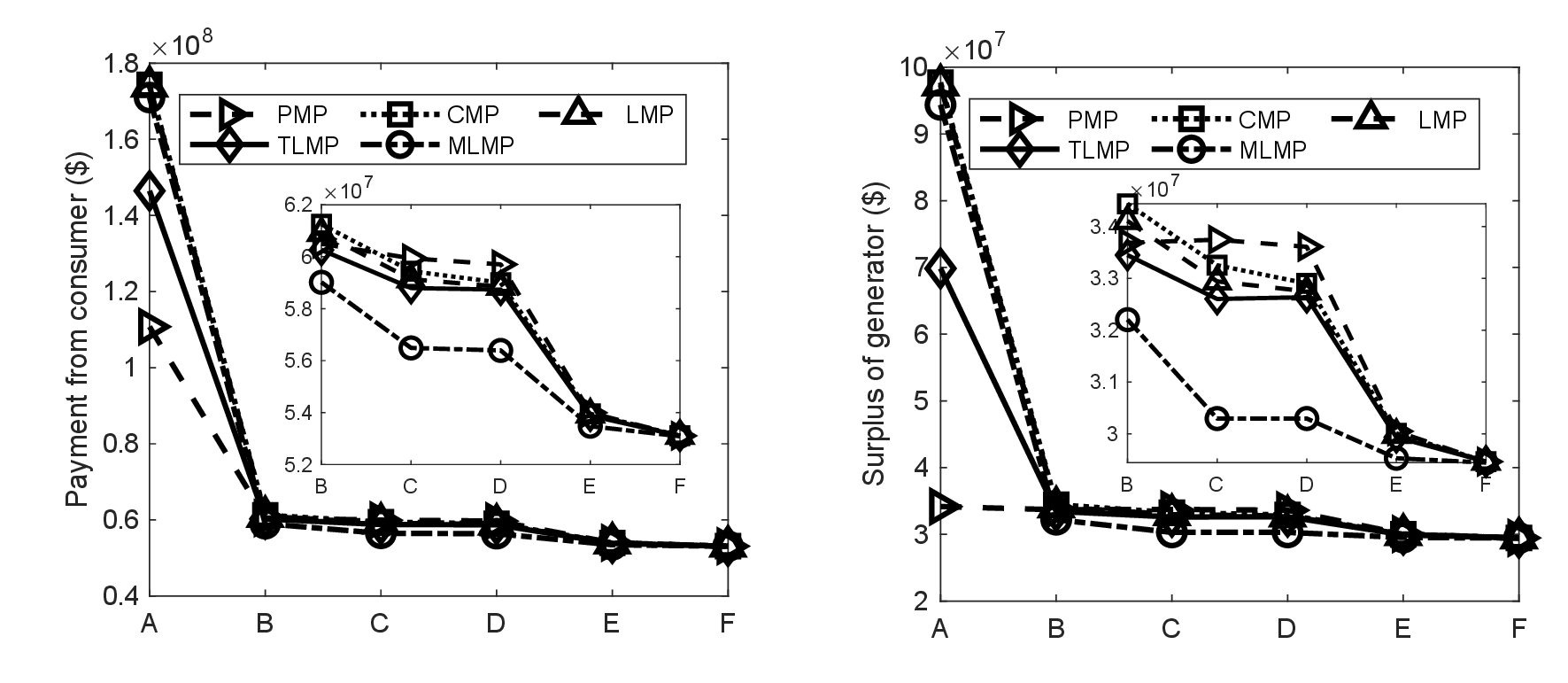}\\
\end{psfrags}
\vspace{-1em}\caption{\small  Left: consumer payment vs ramping limits. Right: surplus of generators vs ramping limits.}
 \label{fig:ISONE_ConsumerPayvsRamp}
\end{figure}

\begin{figure}[h]
\center
\begin{psfrags}
\scalefig{0.4}\epsfbox{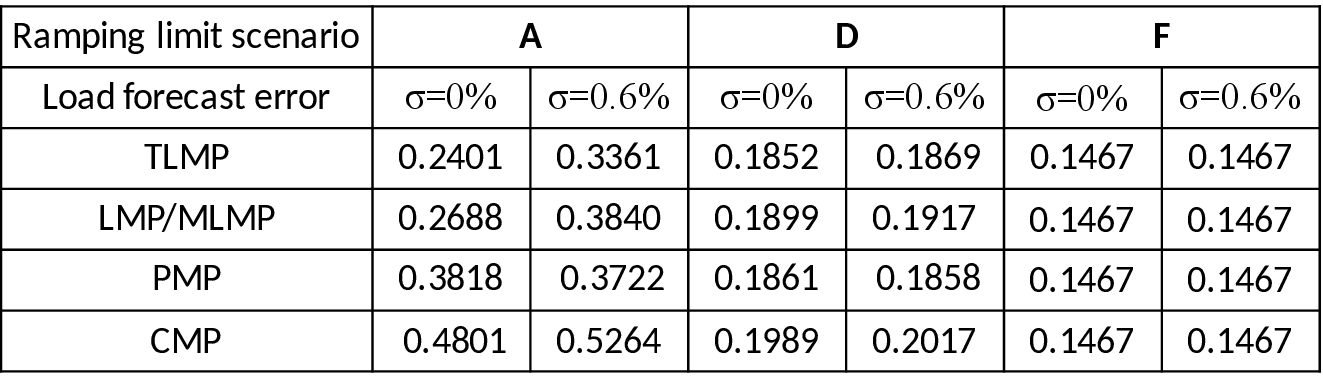}
\scalefig{0.5}\epsfbox{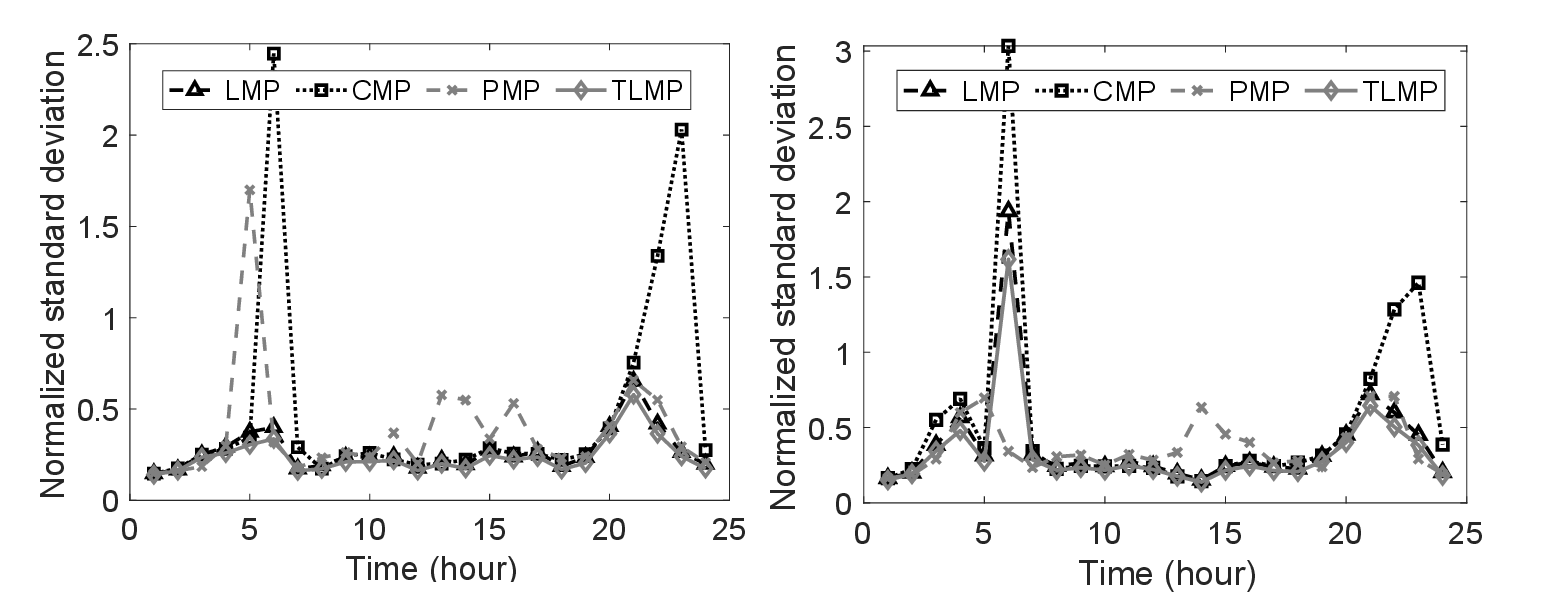}\\
\end{psfrags}
\vspace{-0.1em}\caption{\small Average ratio of normalized standard deviation  of hourly prices. Top: normalized standard deviation under different ramping scenarios and standard deviation of forecasting errors.  Bottom: normalized standard deviation at different hours with $\sigma=0\%$ (left) and $\sigma=6\%$  (right).}
 \label{fig:ISONE_PriceSTD}
\end{figure}

\subsubsection{Revenue adequacy of ISO}
Fig.~\ref{fig:ISONE_ISOsurplusvsRamp}  shows the merchandising surplus of ISO with (left) and without (right)  congestion rent.  Once congestion rent is removed, TLMP was the only one with positive surplus over the entire ramping path with the highest surplus when ramp limits were tight (scenario A and B). The order of the rest of benchmarks in this larger network was not entirely consistent with the three generator case.  Most noticeably was that PMP had the least surplus for the tight ramp case (A) and the highest for the less ramp-limited scenarios (C and D).

\subsubsection{Consumer payments and generator profits}
Fig~\ref{fig:ISONE_ConsumerPayvsRamp} shows the consumer payments (left) and generator profits (right)  under the assumption that the operator charges its shortfall (and returns its profit) to consumers.

As expected, the consumer payment and generator revenue were strongly correlated.  Unlike previous small case studies, consumer payments and generator profit under PMP  for the 76 generator case were the lowest for scenario A and the highest for scenarios  C and D.  MLMP had the lowest total consumer payment and generator profit for scenarios B, C, D, and E.

\subsubsection{Price volatility}
The price volatility was evaluated under different ramping constraints and load forecast errors with network congestion considered. Fig.~\ref{fig:ISONE_PriceSTD}  shows that the average price standard deviation table under strict (A), relaxed (D), and unconstrained (F) ramping limit scenarios. The conclusion here was consistent with previous two small-scale case studies.  For the most part, TLMP appeared to be the least volatile among pricing schemes.

\edoc

\end{document}